\documentclass[prd,twocolumn,lengthcheck,superscriptaddress,showpacs,letterpaper,nofootinbib]{revtex4-1}

\usepackage{color}
\usepackage{graphicx}  
\usepackage{float}
\usepackage{amsmath}
\usepackage{times}
\usepackage{bm}

\begin{document}


\title{Detecting binary neutron star systems with spin in
advanced gravitational-wave detectors.}

\author{Duncan A.\ Brown}
\affiliation{Department of Physics, Syracuse University, Syracuse NY 13244}

\author{Ian Harry}
\affiliation{Department of Physics, Syracuse University, Syracuse NY 13244}

\author{Andrew Lundgren}
\affiliation{Institute for Gravitation and the Cosmos, The Pennsylvania State University, University Park, PA 16802}
\affiliation{Albert-Einstein-Institut, Callinstr. 38, 30167 Hannover, Germany}

\author{Alexander H.\ Nitz}
\affiliation{Department of Physics, Syracuse University, Syracuse NY 13244}


\begin{abstract}
The detection of gravitational waves from binary neutron stars
is a major goal of the gravitational-wave observatories Advanced LIGO and
Advanced Virgo. Previous searches for binary neutron stars with LIGO and Virgo
neglected the component stars' angular momentum (spin).  We demonstrate that
neglecting spin in matched-filter searches causes advanced detectors
to lose more than 3\% of the possible signal-to-noise ratio for 59\% (6\%) of
sources, assuming that neutron star dimensionless spins, $c\mathbf{J}/GM^2$, are uniformly distributed
with magnitudes between $0$ and $0.4$ $(0.05)$ and that the neutron stars
have isotropically distributed spin orientations.
We present a new method for constructing template banks for gravitational
wave searches for systems with spin. We present a new metric in a parameter
space in which the template placement metric is globally flat.
This new method can create template banks of signals with
non-zero spins that are (anti-)aligned with the orbital angular momentum.  We show that this search loses more than
3\% of the maximium signal-to-noise for only 9\% (0.2\%) of BNS sources with dimensionless spins between $0$ and $0.4$ $(0.05)$ and isotropic spin orientations. Use of this
template bank will prevent selection bias in gravitational-wave searches and
allow a more accurate exploration of the distribution of spins in binary
neutron stars.
\end{abstract}

\maketitle

\section{Introduction}

The second-generation gravitational wave detectors Advanced LIGO (aLIGO) and
Advanced Virgo (AdV)~\cite{Harry:2010zz, aVirgo} are expected to begin
observations in 2015, and to reach full sensitivity by 2018-19. These detectors
will observe a volume of the universe more than a thousand times greater than
first-generation detectors and establish the new field of gravitational-wave
astronomy. Estimated detection rates for aLIGO and AdV suggest that binary
neutron stars (BNS) will be the most numerous source detected, with plausible
rates of $\sim 40/\mathrm{yr}$~\cite{Abadie:2010cf}.
Gravitational wave
observations of BNS systems will allow measurement of the properties of
neutron stars and allow us to explore the processes of stellar evolution.

The gravitational waves that advanced detectors will observe from inspiralling BNS systems
are well described by post-Newtonian theory~\cite{Blanchet:2006zz}.
As the neutron stars orbit each other, they lose energy to gravitational waves
causing them to spiral together and eventually merge.
If the
angular momentum (spin) of the component neutron stars is zero, the gravitational
waveform emitted depends at leading order on the chirp mass of the binary
$\mathcal{M} = \left(m_1 m_2\right)^{3/5}/\left( m_1 +
m_2\right)^{1/5}$~\cite{Peters:1963ux}, where $m_1,m_2$ are the component masses
of the two neutron stars, and at higher order on the symmetric
mass ratio $\eta = m_1 m_2 /
(m_1+m_2)^2$~\cite{Blanchet:1995fg,Blanchet:1995ez,BIWW96,Wi93,BFIJ02,Blanchet:2004ek}.
If the neutron stars are rotating, 
coupling between the neutron stars' spin $\bm{S}_{1,2}$ and the
orbital angular momentum $\bm{L}$ of the binary will affect the dynamics of BNS
mergers~\cite{Kidder:1992fr,Apostolatos:1994mx,Kidder:1995zr,Blanchet:2006gy}.  
We measure the neutron stars' spin using the dimensionless parameter
$\bm{\chi}_{1,2} = {\bm{S}_{1,2}}/{m_{1,2}^2}$.

The maximum spin value for a wide class of neutron star equations of state is
$\chi \equiv \left| \bm{\chi} \right| \sim 0.7$~\cite{Lo:2010bj}. However, the spins of neutron stars in BNS
systems is likely to be smaller than this limit. The spin period at the birth
of a neutron star is thought to be in the range
$10$--$140$~ms~\cite{Lorimer:2008se,Mandel:2009nx}. During the evolution of
the binary, accretion may increase the spin of one of the
stars~\cite{Bildsten:1997vw}, however neutron stars are unlikely to have
periods less than 1~ms~\cite{Chakrabarty:2008gz}, corresponding to a
dimensionless spin of $\chi \sim 0.4$.  The period of the fastest known pulsar
in a double neutron star system, J0737--3039A, is
$22.70$~ms~\cite{Burgay:2003jj}, corresponding to a spin of only $\chi \sim
0.05$. In this paper, we therefore consider two populations of neutron star
binaries: the first has spins uniformly distributed from $\chi = 0$ to $0.4$,
the second, a sub-set of this, has spins between $0$ and $0.05$.  This extended spin
distribution allows for the possibility of serendipitous discovery of BNS
systems in globular clusters, where the evolutionary paths may be different
than that in field binaries~\cite{Grindlay:2005ym}. Since supernova kicks may
cause the direction of the neutron star's angular momentum to be misaligned
with the orbital angular momentum of the binary~\cite{Farr:2011gs}, or the
binaries may be formed by direct capture, we consider  a population of
binaries with an isotropic spin distribution.

Searches
for binary neutron star systems in gravitational-wave detectors use
template-based searches~\cite{Allen:2005fk}. Data from the detector is
correlated against a bank of known template waveforms, which cover the space
of parameters searched over~\cite{OwenSathyaprakash98}. The template bank is
constructed so that it covers the parameter space of interest
so that any signal in this region will lose no more
than $3\%$ of the signal-to-noise ratio obtained by an exactly matching
template. Alternative search methods have been proposed~\cite{Marion:2004,Cannon:2010qh},
however these still require the construction of a template bank to perform the
search.
The effect of spin-orbit and spin-spin interactions were neglected in previous
BNS searches~\cite{Abadie:2011nz}, as they do not have a significant effect on
the $\sim 1600$ gravitational wave cycles in the 40--2000~Hz sensitive band of
first-generation detectors~\cite{Apostolatos:1996rf}. However, aLIGO and AdV
will be sensitive to gravitational-wave frequencies between 10--2000Hz,
increasing the number of cycles in band by an order of magnitude.
Initial studies have demonstrated that over this band, the small secular
effects produced by spin-orbit and spin-spin coupling will have a significant
effect on the detectability of BNS systems with non-trivial component
spins~\cite{Ajith:2011ec}. However, the current geometric method for placing
BNS templates~\cite{Bank06} does not incorporate spin. While numerical
(stochastic) methods could be used to include spin, these require
substantially more templates than a comparable geometric
approach~\cite{Harry:2009ea}. 

We present a new geometric algorithm for placing templates for BNS systems
with spin, which has a significantly higher sensitivity than previous
searches.  Our new algorithm constructs a metric on the parameter space using
the various coefficients of the TaylorF2 expansion of the orbital phase as
coordinates. In such a coordinate system the parameter space metric is
globally flat, therefore we can transform into a Euclidean coordinate system.
Finally, our method uses a Principal Coordinate Analysis to identify a two
dimensional manifold that can be used to cover the aligned spin BNS parameter
space using existing two dimensional lattice placement algorithms.

To demonstrate our new method, we first perform a systematic evaluation of the ability
of a search that neglects spin to detect gravitational waves for BNS in aLIGO
and AdV.  We show that this search will lose more than $3\%$ of the
matched filter signal-to-noise ratio for 59\% (6\%) of signals if it is used to search
for BNS systems with spins uniformly distributed between $0 \le \chi_{1,2} \le
0.4 (0.05)$; this is unsatisfactory over a
large region of the signal parameter space. We show that by considering BNS
systems where the spin of the neutron stars are aligned with the orbital
angular momentum (i.e. the binary is not precessing), we can create a
two-dimensional template bank that is efficient at detecting spin-aligned BNS
signals. Finally we demonstrate that this bank is sufficient to detect signals
from generic spinning, precessing binaries in aLIGO and AdV. The spin-aligned
bank loses more than $3\%$ of the signal-to-noise ratio for only 9\% (0.2\%)
of signals, sufficient to construct a sensitive and unbiased search for BNS
systems in aLIGO and AdV.

\section{BNS Search Sensitivity}
\label{ssec:nonspin_performance}
\label{sec:spin_import}

We quantify the performance of templated matched-filter searches by the
fitting factor (FF) of the search~\cite{Apostolatos:1995pj}.  The fitting
factor is the fraction of the signal-to-noise ratio that would be recovered
when matching a given signal with the best matching waveform in the template
bank. We first define the overlap between two templates $h_1$ and $h_2$ as
\begin{equation}
\mathcal{O}(h_1,h_2) = (\hat{h}_1|\hat{h}_2) = \dfrac{(h_1|h_2)}{\sqrt{(h_1|h_1)(h_2|h_2)}}.
\end{equation}
which is defined in terms of the noise-weighted inner product~\cite{CF94}
\begin{equation}
(h_1|h_2) = 4 \, \mathrm{Re} \int^{\infty}_0\dfrac{\tilde{h}_1(f)\tilde{h}_2^*(f)}{S_n(f)} df.
\end{equation}
This overlap is the fraction of signal power that would be recovered by
searching for the signal $h_1$ using a matched filter constructed from $h_2$.
Maximizing the overlap over the time of arrival and waveform phase yields the
match
\begin{equation}
\mathcal{M}(h_1,h_2) = \underset{\phi_c,t_c}{\max}(\hat{h}_1|\hat{h}_2(\phi_c,t_c)).
\end{equation}
The mismatch, $1 - \mathcal{M}$, is the fraction of the optimal
signal-to-noise ratio that is lost when searching for a signal $h_1$ with
a template waveform $h_2$. 


When searching for BNSs, we do not know the exact physical parameters of the
system. We assume that the masses of the neutron stars lie between $1$ and
$3\, M_\odot$ and construct a bank of waveform templates $\{h_b\}$ to span this
region of the mass parameter space. To measure the sensitivity of this bank to
a gravitational waveform $h_s$ with unknown parameters, we compute the fitting
factor
\begin{equation}
\textrm{FF}(h_s) = \max_{h \in \{h_b\}} \mathcal{M}(h_s,h),
\end{equation}
where we have maximized the match over all the templates in the bank.  In
searches for gravitational waves using LIGO and Virgo, the bank is constructed
such that the fitting factor for any signal in the target parameter space will
never be less than 0.97. At least one of the templates in the bank must have a
maximized overlap of 0.97 (or more) with the signal. This value is chosen to
correspond to an event rate loss of no more than 10\% of possible sources
within the range of the detectors~\cite{Cutler:1992tc}. In this paper, we use
a fitting factor of 0.97 to construct search template banks.

We now test whether a bank of templates that does not model the effect of spin
is sufficient to detect generic, spinning BNS sources in aLIGO and AdV. We
create a bank of non-spinning templates that would recover any
non-spinning BNS system with a fitting factor greater than 0.97.  This bank is
constructed using TaylorF2 waveforms, which are constructed using the stationary
phase approximation to the gravitational-wave phasing accurate to 3.5
post-Newtonian (PN) order~\cite{DPK99,Blanchet:2006zz}. To create a
bank of these waveforms we use the hexagonal-placement method defined in
\cite{Cokelaer:2007kx}, which was used in the majority of previous searches in
LIGO and Virgo~\cite{Abbott:2009tt,Abbott:2009qj,Abadie:2010yba}. This
template bank is placed using the metric given in \cite{OwenSathyaprakash98},
which is valid, by construction, for templates at 2PN order. 
Our signal waveforms are constructed using the SpinTaylorT4
waveform \cite{BCV03b}, a time-domain waveform accurate to 3.5PN order
in the orbital phase which includes the leading order spin-orbit, spin-spin,
and precessional modulation effects and implemented in the LSC Algorithm Library Suite
\cite{lalsuite}. We first confirm that although the
bank is constructed at 2PN order, it yields fitting factors greater than 0.97
for both the TaylorF2 and SpinTaylorT4 non-spinning waveforms at 3.5PN order.
To simulate a population of spinning BNS sources, we generate 100,000 signals
with component masses uniformly distributed between 1 and 3 $M_{\odot}$ and
dimensionless spin magnitudes uniformly distributed between 0 and 0.4. The orientation
of the spin, the orientation of the orbital angular momentum, and the sky
location are isotropically distributed.  To model the sensitivity of a second
generation gravitational wave interferometer, we use the aLIGO zero-detuned,
high-power sensitivity curve \cite{aLIGOSensCurves}. For our simulations, we
use a lower frequency cutoff of 15Hz.

We note that for non-precessing systems
the fitting factor is independent of the detector alignment and location; however
this statement is not true for precessing systems. For such systems, however,
the distribution of fitting factors over
a population of sources will be independent of the detector alignment
and location. Therefore, for this study we calculate the fitting-factor for a single
detector with an arbitrary location and position.

\begin{figure}
\includegraphics[width=0.45\textwidth]{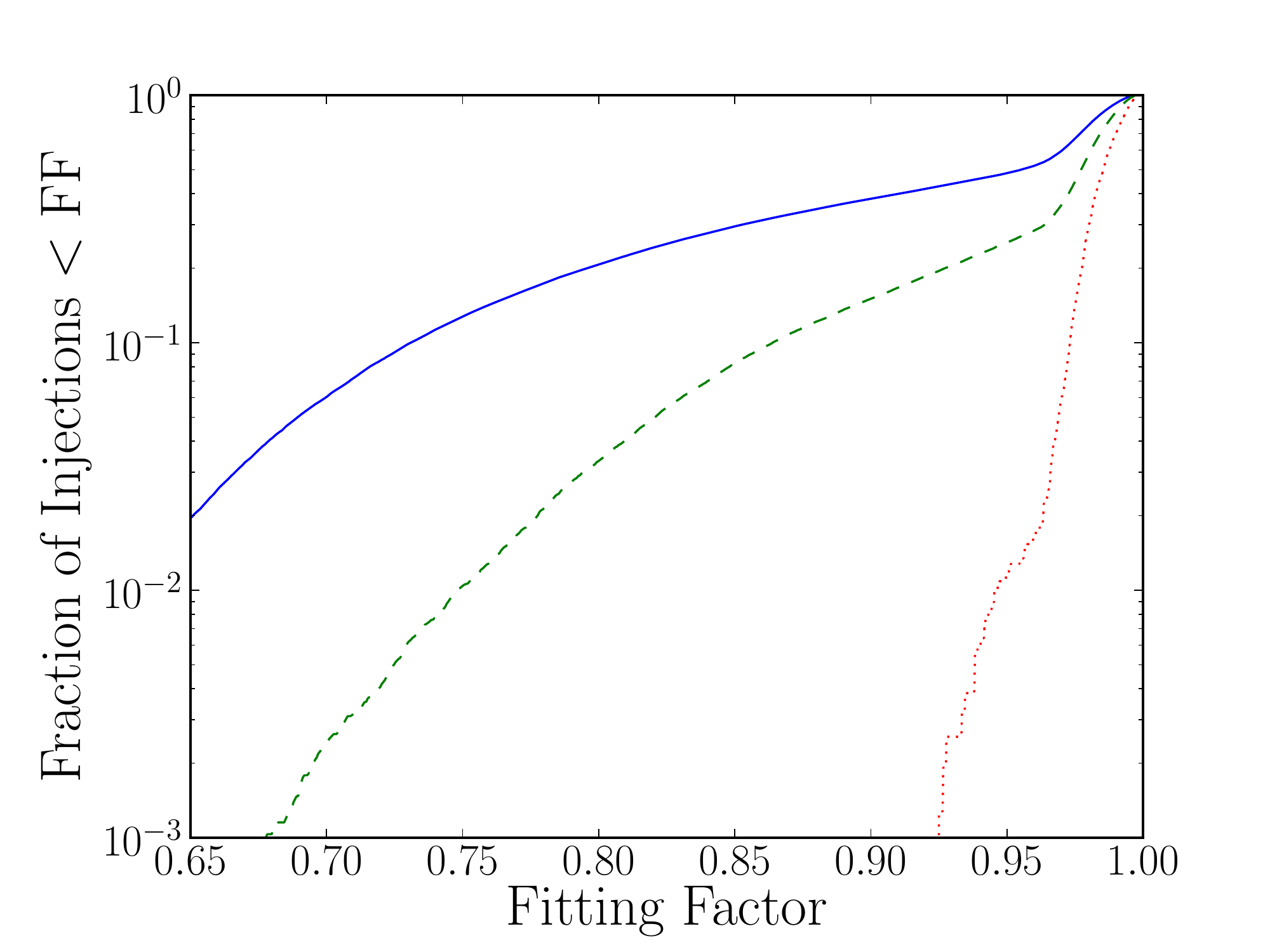}
\caption{\label{fig:no_spin_cover} The distribution of fitting factors obtained by searching
for the precessing BNS systems described in section \ref{ssec:nonspin_performance}
with component spins up to 0.4 (blue solid line), 0.2 (green dashed line), and 0.05 (red dotted line) using the non-spinning
BNS template bank described in section \ref{ssec:nonspin_performance} and the advanced LIGO, zero-detuned,
high-power PSD with a 15Hz lower frequency cutoff.}
\end{figure}

In Fig.~\ref{fig:no_spin_cover} we show the distribution of fitting factors
obtained when searching for our population of BNS sources with the
non-spinning template bank. We see that 59\% of signals were recovered with a
fitting factor less than 0.97.  If the maximum spin magnitude is restricted to
0.05, we find that 6\% of signals are recovered with a FF less
than 0.97.  If BNS systems do exist with spin magnitudes up to 0.4, a template
bank that captures the effects of spin will be required to maximize the number
of BNS detections.  Detection efficiency will be greatly reduced by using a
template bank that only contains waveforms with no spin effects.  Even under
the assumption that component spins in BNS systems will be no greater
than 0.05, detection efficiency will be decreased if the effect of spin on the
signal waveform is ignored.

\section{A template placement algorithm for aligned-spin BNS templates}
\label{sec:param_space}

As we have demonstrated in the previous section, there is a substantial region
of the BNS parameter space where a significant loss in signal-to-noise ratio
would be encountered when searching for astrophysically plausible, spinning
BNS systems with non-spinning templates. It has been suggested that using BNS
templates where the spins of the system are aligned with the orbital angular
momentum is sufficient for detecting generic BNS systems with second-generation
detectors~\cite{Ajith:2011ec} using TaylorF2 templates that incorporate the
leading order spin-orbit and spin-spin corrections~\cite{PW95}. 

In this section we use these spin-aligned waveforms to construct a template
bank that attempts to cover the full space of astrophysically plausible BNS
spin configurations. This template bank should contain as few templates as
possible, while still being able to detect any BNS system that might be
observed with aLIGO and AdV. To achieve this, it is important to assess the
``effective dimension'' of the space, which is defined as the number of orthogonal
directions
over which template waveforms need to placed in order to cover the full physically
possible parameter range. 
We demonstrate that the effective dimension
of this parameter space is only two dimensional. For BNS systems in aLIGO and AdV
the extent of the physical parameter space in the remaining directions is smaller
than the coverage radius of a template and can be neglected.

As the effective dimension of the space is two-dimensional, a
hexagonal placement algorithm, similar to that used in previous searches of
LIGO and Virgo data, could be employed to cover the space. This allows our new
method to be incorporated into existing search pipelines in a straightforward
way.


Since BNS systems coalesce at $\sim 1500$~Hz, significantly higher than the
most sensitive band of the detectors, the waveform will be dominated by the
inspiral part of the signal~\cite{Buonanno:2009zt}.  The effect of component
spin on BNS inspiral waveforms has been well explored in the
literature~\cite{Apostolatos:1994mx,Kidder:1992fr,Kidder:1995zr,BCV03b}).
For spin-aligned (i.e. non-precessing) waveforms, the dominant effects of
component spin are spin-orbit coupling, which enters the waveform phasing at
1.5PN order, and spin1-spin2 coupling, which enters the waveform phasing at
2PN order.  Other spin-related corrections to the PN phasing have been
computed \cite{Mikoczi:2005dn,Arun:2008kb}, however, in this work we mainly restrict
to only the two dominant terms. The methods described here are easily
extendable to include additional spin correction terms and this does not
significantly change our results, as we demonstrate at the end of this section.

\begin{widetext}
To construct a bank to search for generic BNS signals, we use TaylorF2
waveforms accurate to 3.5PN order in orbital phase and including the
leading order spin-orbit and spin-spin terms given by~\cite{PW95,Buonanno:2009zt}
\begin{equation}
 \tilde h(f) = A(f;\theta_x) e^{i \Psi(f;\lambda_i)}\
\end{equation}
where $\theta_x$ describe the various orientation angles that only affect the amplitude and overall phase of the observed gravitational waveform~\cite{Allen:2005fk}. The phase
$\Psi$ is given by
\begin{equation} \label{eq:phase_exp}
\Psi = 2 \pi f_0 x t_c - \phi_c + \lambda_0 x^{-5/3} + \lambda_2 x^{-1} + \lambda_3 x^{-2/3} + \lambda_4 x^{-1/3} 
 + \lambda_{5L} \log(x) + \lambda_6 x^{1/3} + \lambda_{6L} \log(x) x^{1/3} + \lambda_{7} x^{2/3},
\end{equation}
where $f$ is the frequency, $f_0$ is a fiducial frequency, $x = f/f_0$, $t_c$ is the coalescence time,
$\phi_c$ is a constant phase offset. The PN phasing terms are
\begin{eqnarray}
\label{eq:lambdas35}
\lambda_0 &=& \frac{3}{128} (\pi \mathcal{M} f_0)^{-5/3}, \\
\lambda_2 &=& \frac{5}{96 \eta^{2/5}} \left( \frac{743}{336} + \frac{11}{4} \eta \right) (\pi \mathcal{M} f_0)^{-1}, \\
\lambda_3 &=& -\frac{3 \pi}{8 \eta^{3/5}} \left( 1- \frac{1}{4 \pi} \beta \right) (\pi \mathcal{M} f_0)^{-2/3}, 
\end{eqnarray}
\begin{eqnarray}
\lambda_4 &=& \frac{15}{64 \eta^{4/5}} \left( \frac{3058673}{1016064} + \frac{5429}{1008} \eta + \frac{617}{144} \eta^2 - \sigma \right) (\pi \mathcal{M} f_0)^{-1/3} \\
\lambda_{5L} &=& \frac{3}{128 \eta} \left(\frac{38645 \pi}{756} - \frac{65 \pi}{9}\eta\right) \\
\lambda_6 &=& \frac{3}{128\eta^{6/5}} \biggl(\frac{11583231236531}{4694215680} - \frac{640 \pi^2}{3} 
          - \frac{6848}{21} \left( \gamma_E + \log 4 - \frac{1}{5} \log \eta + \frac{1}{3} \log (\pi \mathcal{M} f_0) \right) \nonumber \\
          &-& \frac{15737765635}{3048192}\eta + \frac{2255 \pi^2}{12}\eta + \frac{76055}{1728}\eta^2
          - \frac{127825}{1296}\eta^3 \biggr) (\pi \mathcal{M} f_0)^{1/3} \\
\lambda_{6L} &=& -\frac{1}{128 \eta^{6/5}}\frac{6848}{21} (\pi \mathcal{M} f_0)^{1/3} \\
\lambda_{7} &=& \frac{3}{128\eta^{7/5}} \left(\frac{77096675 \pi}{254016} + \frac{378515 \pi}{1512}\eta - \frac{74045 \pi}{756}\eta^2\right) (\pi \mathcal{M} f_0)^{2/3},
\end{eqnarray}
where $\gamma_E$ is the Euler gamma constant, $\beta$ (the dominant spin-orbit coupling term)
and $\sigma$ (the dominant spin-spin coupling term) are given by
\begin{eqnarray}
\beta &=& \frac{1}{12} \sum_{i=1}^2 \left[ 113 \left(\frac{m_i}{m_1 + m_2}\right)^2 + 75 \eta \right]
\bm{\hat{L}} \cdot \bm{\chi}_i \\
\sigma &=& \frac{\eta}{48} \left( -247 \bm{\chi}_1 \cdot \bm{\chi}_2 + 721 \bm{\hat{L}}
\cdot \bm{\chi}_1 \bm{\hat{L}} \cdot \bm{\chi}_2 \right).
\end{eqnarray}
and $\bm{\hat{L}}$ is the unit vector in the direction of the orbital angular momentum.
Note that above we have omitted the $\lambda_5$ term, as it has no dependance on frequency
and is therefore included in the constant phase offset, $\phi_c$.
\end{widetext}

Our goal is to construct a template bank containing the minimum number of
waveforms for which any plausible BNS signal has a FF of 0.97 or higher.
To place a template bank, we follow the method of Owen~\cite{Owen96}. We first
construct a metric on the waveform parameter space that describes the
mismatch between infinitesimally separated points,
\begin{equation}
\mathcal{O}(h(\bm{\theta}),h(\bm{\theta}+\delta\bm{\theta})) =
  1 - \sum_{ij} g_{ij}(\bm{\theta}) \,\delta\theta^i \,\delta\theta^j,
\end{equation}
with the metric given by,
\begin{equation}
\label{eq:cbc_metric}
g_{ij}(\boldsymbol{\theta}) = - \frac{1}{2} \frac{\partial^2 \mathcal{O}}{\partial \delta\theta^i \partial
\delta\theta^j} = \left(\frac{\partial h(\boldsymbol{\theta})}{\partial \theta^i} \bigg|
\frac{\partial h(\boldsymbol{\theta})}{\partial \theta^j}\right)
\end{equation}
and where $\bm{\theta}$ describes the parameters of the signal, in this case the masses and
the spins. 

This metric is used to approximate the mismatch in the neighborhood
of any point. When doing this care must be taken to choose a ``good'' set
of coordinates where extrinsic curvature is minimized. If a ``bad'' set of
coordinates is chosen, the region in which this approximation can be used will
be very small. To minimize this issue when placing the two-dimensional
non-spinning bank, the masses $m_1, m_2$ are transformed into the ``chirp
times'' $\tau_0, \tau_3$~\cite{OwenSathyaprakash98}. In this coordinate
system, ellipses are constructed that describe fitting factors greater than 0.97
around a point and hexagonal placement is used to efficiently tile the space to
achieve the desired minimal match~\cite{Bank06}.

To construct our new bank, we treat the six $\lambda_i$ and two $\lambda_{i L}$ components, given in
Eq.~(\ref{eq:lambdas35}), as eight independent parameters, as in \cite{Pai:2012mv}.
The range of
possible physical values will trace out a four-dimensional manifold in the
eight dimensional parameter space given by the $\lambda_\alpha$, where $\alpha$ is an index that takes both $i$ and $i L$ values. We will demonstrate
that this eight-dimensional parameter space allows us to construct a metric
without intrinsic curvature.

As shown in \cite{Owen96} it is possible to evaluate the derivative in (\ref{eq:cbc_metric}),
maximizing over the phase, $\phi_C$, to give the metric in terms of a 9 dimensional space:
\begin{equation}
\gamma_{\alpha \beta} = \frac{1}{2} \left( \mathcal{J}[\psi_\alpha \psi_\beta] - \mathcal{J}[\psi_\alpha] \mathcal{J}[\psi_\beta] \right) ~.
\end{equation}
In this expression $\psi_{\alpha}$ is given by
\begin{eqnarray}
\psi_0 &=& \frac{\partial \Psi}{\partial t_c} = 2 \pi f_0 x \\
\psi_i &=& \frac{\partial \Psi}{\partial \lambda_i} = x^{(i-5)/3} \\
\psi_{i L} &=& \frac{\partial \Psi}{\partial \lambda_{i L}} = x^{(i-5)/3} \log(x)
\end{eqnarray}
and $\mathcal{J}$ is the moment functional of the noise PSD \cite{PW95,Owen96}
\begin{equation}
\mathcal{J}[ a(x) ] = \frac{1}{I(7)} \int_{x_L}^{x_U} \frac{a(x) x^{-7/3}}{S_h(x f_0)} \mathrm{d}x ~,
\end{equation}
where
\begin{equation}
I(q) \equiv \int_{x_L}^{x_U} \frac{x^{-q/3}}{S_h(x f_0)} \mathrm{d}x
\end{equation}
and $x_U$ and $x_L$ correspond to the lower and upper bounds of frequency in the integral. Unless stated
otherwise we use $f_L = x_L f_0 = 15\mathrm{Hz}$ for the aLIGO PSD and choose $2000$Hz
for the upper frequency cutoff, $f_U = x_U f_0$. While it is unphysical
to use the same upper frequency cutoff for all systems, especially as we are not including a merger in our
waveform model, it is necessary to make this assumption to ensure that our metric will be flat. For BNS systems
this approximation is fair to use
as such systems will merge at frequencies that are outside the sensitive range of the advanced detectors and
thus our calculation of signal power is not affected by assuming that all BNS systems merge abruptly at $2000$Hz.
This approach
was also used in \cite{Bank06} for computational efficiency.

Following \cite{Owen96} we can then maximize this expression over $t_C$ to give the metric in terms of the eight $\lambda_\alpha$ 
\begin{equation}
 g_{\alpha \beta} = \gamma_{\alpha \beta} - \frac{\gamma_{0 \alpha} \gamma_{0 \beta}}{\gamma_{0 0}}.
\end{equation}
It is worth highlighting that the parameter space
metric $g_{\alpha \beta}$, in the $\lambda_\alpha$ coordinate system, has no dependence
on the values of $\lambda_\alpha$. In other
words, the parameter space is globally flat in this eight-dimensional
parameter space.

Although this eight-dimensional metric is globally flat, we have increased the
dimensionality of the physical waveform space by a factor of two.  However, we
can transform this metric to a new coordinate system that will allow us to
assess the effective dimensionality of the parameter space.
We first rotate and rescale the metric to transform to a Cartesian
coordinate system. We now use indicies $i, j$ to number the remaining eight $\lambda_\alpha$ coordinates. As $g_{ij}$ is a real, symmetric matrix we can use the
eigenvalues and eigenvectors of the metric to rotate into an orthonormal
coordinate system defined by
\begin{equation}
 \mu_i = \sum_{j} \left(V_{ij} \sqrt{E_i} \right) \lambda^{j},
\end{equation}
where $V_{ij}$ describes the eigenvectors of $g_{ij}$ and $E_{i}$ its
corresponding eigenvalues. 
We use the convention that $V_{ij}$ is the $j^{th}$ component of the $i^{th}$
eigenvector, and the eigenvectors are normalized by $V^T V = \mathcal{I}$. 
In this coordinate system, the metric, $g'_{ij}$,
will be the identity matrix. Next, we perform a rotation to align the axis of
the parameter space with the principal components of the physically possible
region of the space. The physically allowed ranges of the masses and spins
cover only a limited region in the parameter space.  The extent of the
physically relevant region of the space in a certain direction may be thin
relative to the desired mismatch. By orienting the coordinate system along the
principal directions we can easily identify any orthogonal directions in which
the physical region is sufficiently thin that we do not need to place
templates in those directions. This will allow us to assess the effective
dimension of the parameter space, or, in other words, how many directions we
need to consider when placing a template bank. Transforming to a Cartesian
coordinate system also helps with template placement, as it is trivial to
place templates using the optimal $A_n^*$ lattice \cite{Conway:1993} in a 2, 3
or 4 dimension Cartesian coordinate system

To perform the second rotation we make use of the fact that in a Cartesian
coordinate system we are free to rotate the coordinates without changing the
form of the metric. We would like to rotate the coordinates so that the
greatest extent of the template bank lies along as few directions as possible.
To accomplish this we first draw many examples of physical parameters of the
masses and spins, and calculate the corresponding values of $\mu_i$ for each
of these points.  We then do a Principal Component Analysis on this dataset,
which amounts to finding the eigenvectors of the covariance matrix from the
set of $\mu_i$.  This produces a rotation into a new set of coordinates given
by
\begin{equation}
 \xi_i = \sum_{j} \left(C_{ij} \mu^{j}\right),
\end{equation}
where $C_{ij}$ contains the eigenvectors of the covariance matrix 
using the same conventions as for $V_{ij}$. 
The rotation of course leaves the metric Cartesian, but now the bank is oriented
along the principal axes and it is much easier to visualize the shape of the
boundaries and determine how to perform the template placement.

\begin{figure}[!ht]
\includegraphics[width=0.44\textwidth]{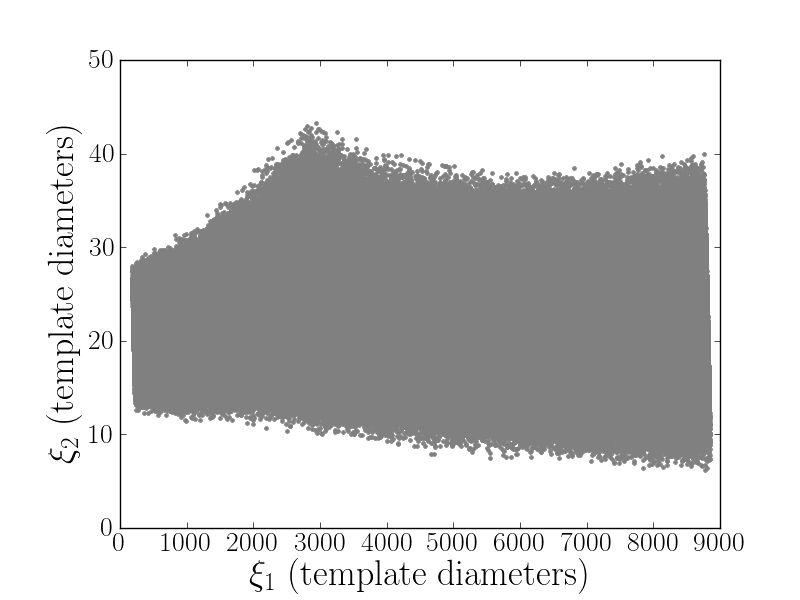}
\includegraphics[width=0.44\textwidth]{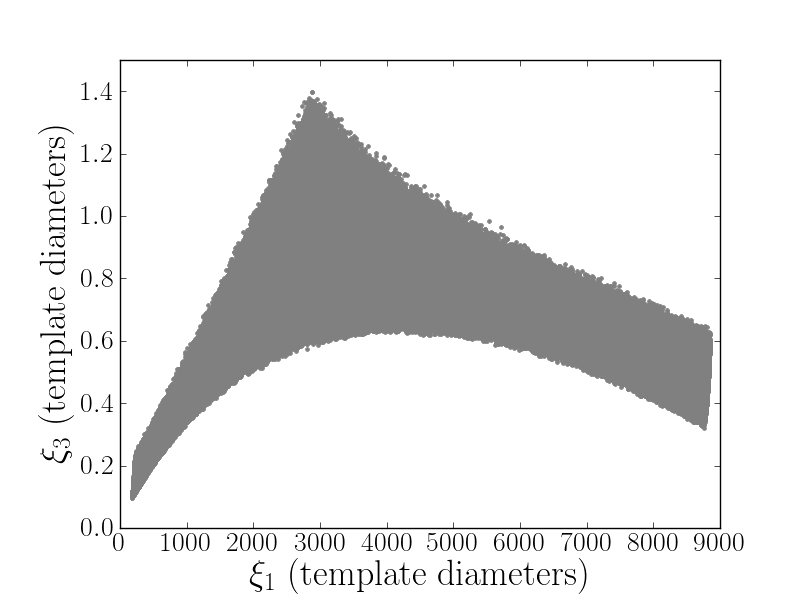}
\includegraphics[width=0.44\textwidth]{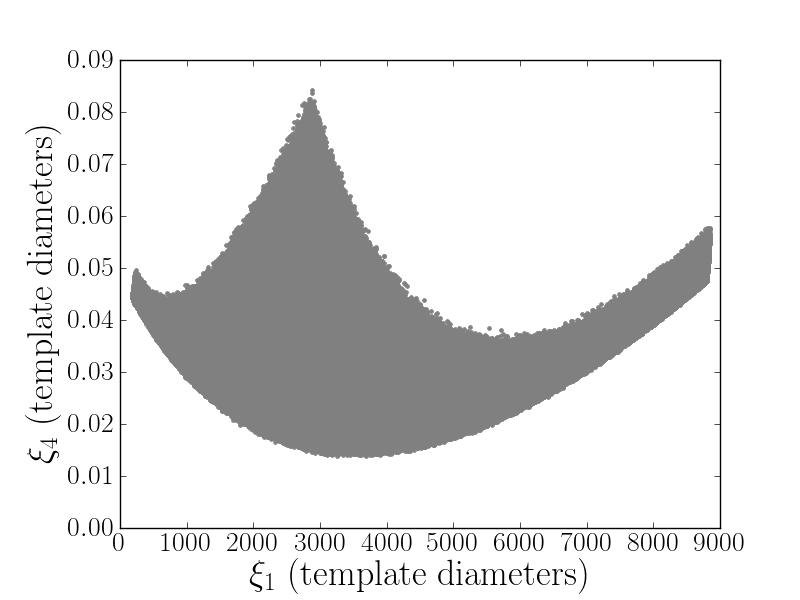}
\caption{\label{fig:param_space_extent} The extent of the binary neutron star, $\chi_i < 0.4$,
parameter space in the $\xi_2$, $\xi_3$ and $\xi_4$
directions, plotted against $\xi_1$. The $\xi_i$ coordinates have been scaled
such that one unit corresponds to the coverage diameter of a template
at 0.97 mismatch. Generated
using the zero-detuned, high-power advanced LIGO sensitivity curve with a 15Hz lower frequency cut off.
}
\end{figure}

We now use this method to construct a template bank where the spin of each
component neutron star is restricted to 0.4. When this metric is constructed
using the aLIGO, zero-detuned high-power noise curve with a lower
frequency cut-off of 15Hz we show that, although many additional templates are
required to cover an aligned spinning parameter space when compared to the non
spinning space, the effective dimension for these BNS systems is still two.

We begin by attempting to visualize the space. We will refer to $\xi_1$ as the
direction along which the parameter space has the biggest extent (the dominant
direction) and $\xi_8$ as the direction with the smallest extent (the
least-dominant direction).  We draw a large set of points, with random values
of masses and spins, and transform these points into the $\xi_i$ coordinate
system. The position of these points is shown in Figure
\ref{fig:param_space_extent}, where we plot the extent of $\xi_{2,3,4}$
against $\xi_1$.

In Figure \ref{fig:param_space_extent} and subsequent plots, we have scaled the $\xi_i$
direction such that one unit corresponds to the coverage diameter of a template
at 0.97 mismatch. Equivalently, we have scaled the directions such
that two points separated by 0.5 units (one template radius) in any direction have a match of 0.97%
\footnote{The unscaled distance between two points with a match of 0.97 would be
$(1 - 0.97)^{0.5} = 0.17$}. We remind that mismatch is proportional to distance squared
and therefore two points separated by one unit would have a match of 0.88

Immediately we notice that the extent along the $\xi_4$
direction is small compared to the diameter of a template. We can also see that
the extent along the $\xi_3$ direction is comparable to a template
diameter, while the $\xi_1$ and $\xi_2$ directions have much larger extents
and clearly need to be gridded over. The extent in the other 4 directions is
smaller than $\xi_4$ and can be completely ignored. This hierarchy of measurable 
parameters may be a generic feature according to the model of \cite{Transtrum:2010zz}.

\begin{figure}
\includegraphics[width=0.45\textwidth]{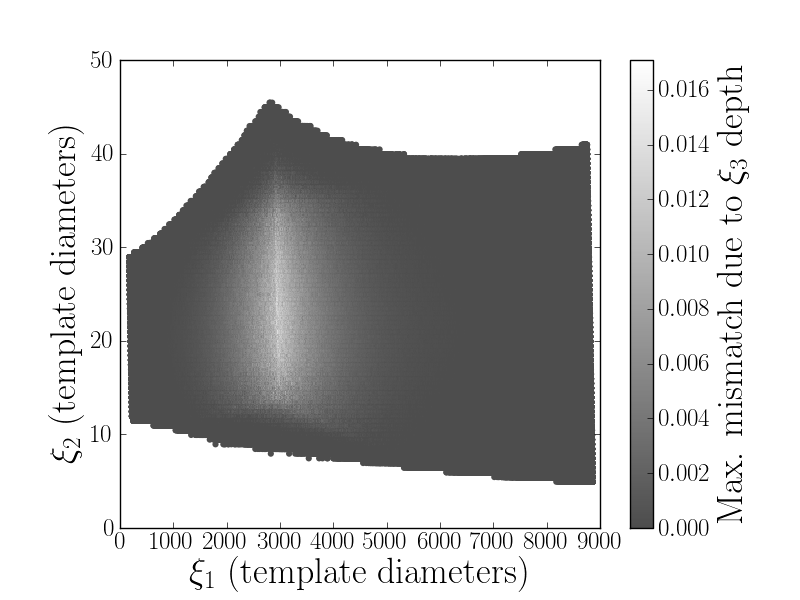}
\includegraphics[width=0.45\textwidth]{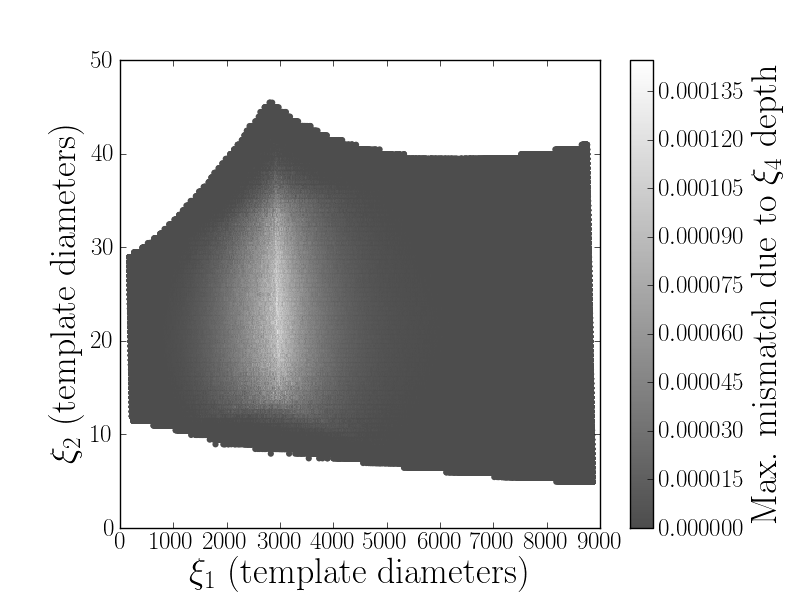}
\caption{\label{fig:param_space_mismatch} The mismatch between the edge and centre of the
physically possible range of $\xi_3$ (top) and $\xi_4$ (bottom) values as a function of
$\xi_1$ and $\xi_2$. The $\xi_i$ coordinates have been scaled
such that one unit corresponds to the coverage diameter of a template
at 0.97 mismatch. Plotted for a binary neutron star parameter space with spins restricted
to 0.4 using the zero-detuned, high-power advanced LIGO sensitivity curve with a 15Hz lower frequency cut off.}
\end{figure}

The plot of $\xi_1$ against $\xi_3$ in Figure \ref{fig:param_space_extent} can be somewhat misleading
as we have projected out the $\xi_2$ direction.
It is more informative to investigate the depth of $\xi_3$ at fixed values of $\xi_1$ and $\xi_2$
and translate this into the maximum mismatch that would be obtained
if one were to assume that there is no width in the third direction.
In Figure \ref{fig:param_space_mismatch} we show the maximum
mismatch between the central and extremal values of the possible range of
$\xi_3$ (and $\xi_4$) as a function of the
two primary directions. This is calculated by binning the points mentioned above into bins in
$\xi_1$ and $\xi_2$, where the bin width is equal to one template radius. We then determine the extremal values of
$\xi_3$ (and $\xi_4$) for the points in each bin. From Figure \ref{fig:param_space_mismatch}
we can see that, while there are small areas
of parameter space where up to a 1.6\% loss in SNR would be incurred from assuming the $\xi_3$
direction had no depth,
most areas of the parameter space are very thin in the $\xi_3$ direction. This figure also helps to reinforce
the fact that the depth in the fourth direction is negligible, as, even in the worst region of the space, no more
than 0.01\% of SNR would be lost by assuming $\xi_4$ had no depth. The depth of the $\xi_{5,8}$ directions
are even smaller than $\xi_4$.

\begin{figure}
\includegraphics[width=0.45\textwidth]{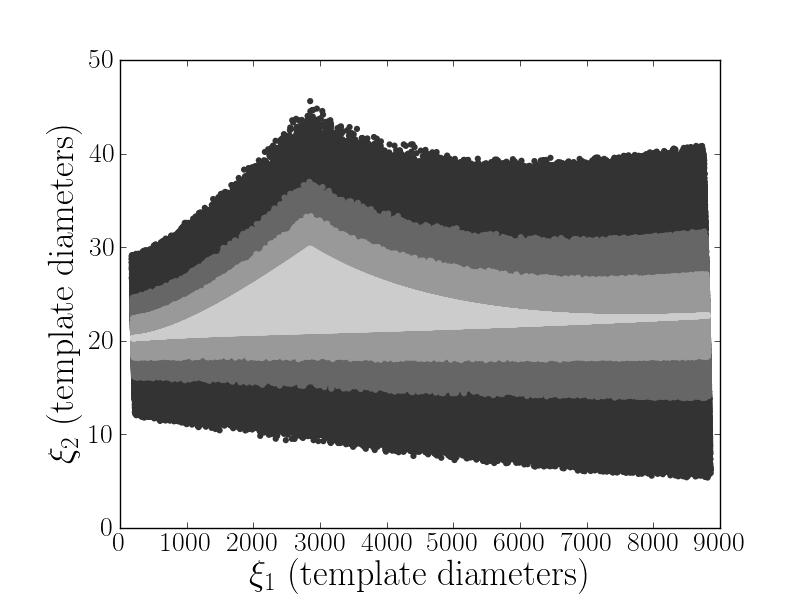}
\caption{\label{fig:param_space_spin} The size of the BNS parameter space as a function of the maximum spin.
The darkest points indicate points with spin on both components constrained to 0.4, then, in order of
increasing lightness, we show points constrained
to a maximal spin of 0.2 and 0.1, finally the lightest points show points with no spin. 
The $\xi_i$ coordinates have been scaled
such that one unit corresponds to the coverage diameter of a template
at 0.97 mismatch.
This plot was generated
using the zero-detuned, high-power aLIGO sensitivity curve with a 15Hz lower frequency cut off.}
\end{figure}

In this coordinate system it is easy to explore how the size of the parameter space
depends on the maximal spins of the component neutron stars. In Figure \ref{fig:param_space_spin} we show the
extent of the physical space for aligned spinning BNS systems, with maximum component
spins of 0.4, 0.2 and 0.1, compared to that
of non-spinning systems. Ignoring any issues related to the depth of the $\xi_3$
direction, one can clearly see that to cover the aligned spin parameter space will require
a great deal more templates than the non spinning parameter space.


From these results we can see that a 2 dimensional template bank would be sufficient to cover the aligned
spin parameter space for BNS systems in the advanced detector era. Specifically, we would advocate
placing a hexagonal lattice in the $\xi_1$, $\xi_2$ coordinates and setting the value of $\xi_{3..8}$
to be the middle of the possible range of those parameters at the given position of $\xi_1$, $\xi_2$.
For the regions of parameter space where the depth of $\xi_3$ is not negligible, one could either ignore
it, understanding that the resulting bank will not have a fitting factor of 0.97 in this region.
Alternatively, one could stack templates in the region where $\xi_3$ is deepest to minimize this effect.


For this work we chose to employ a hexagonal template bank in the $\xi_1$, $\xi_2$ coordinates, stacking the
templates in the $\xi_3$ direction, where necessary, to ensure that the maximum mismatch due to the depth
of $\xi_3$ is no more
than 0.25\%. For an aligned-spin template bank where the spin of each component is restricted to 0.4,
using the advanced LIGO, zero-detuned high-power noise curve with a lower frequency cut-off of 15Hz,
we find that approximately 520,000 templates are required.
Roughly 100,000 of these templates were added by the stacking process.

\begin{figure}
\includegraphics[width=0.45\textwidth]{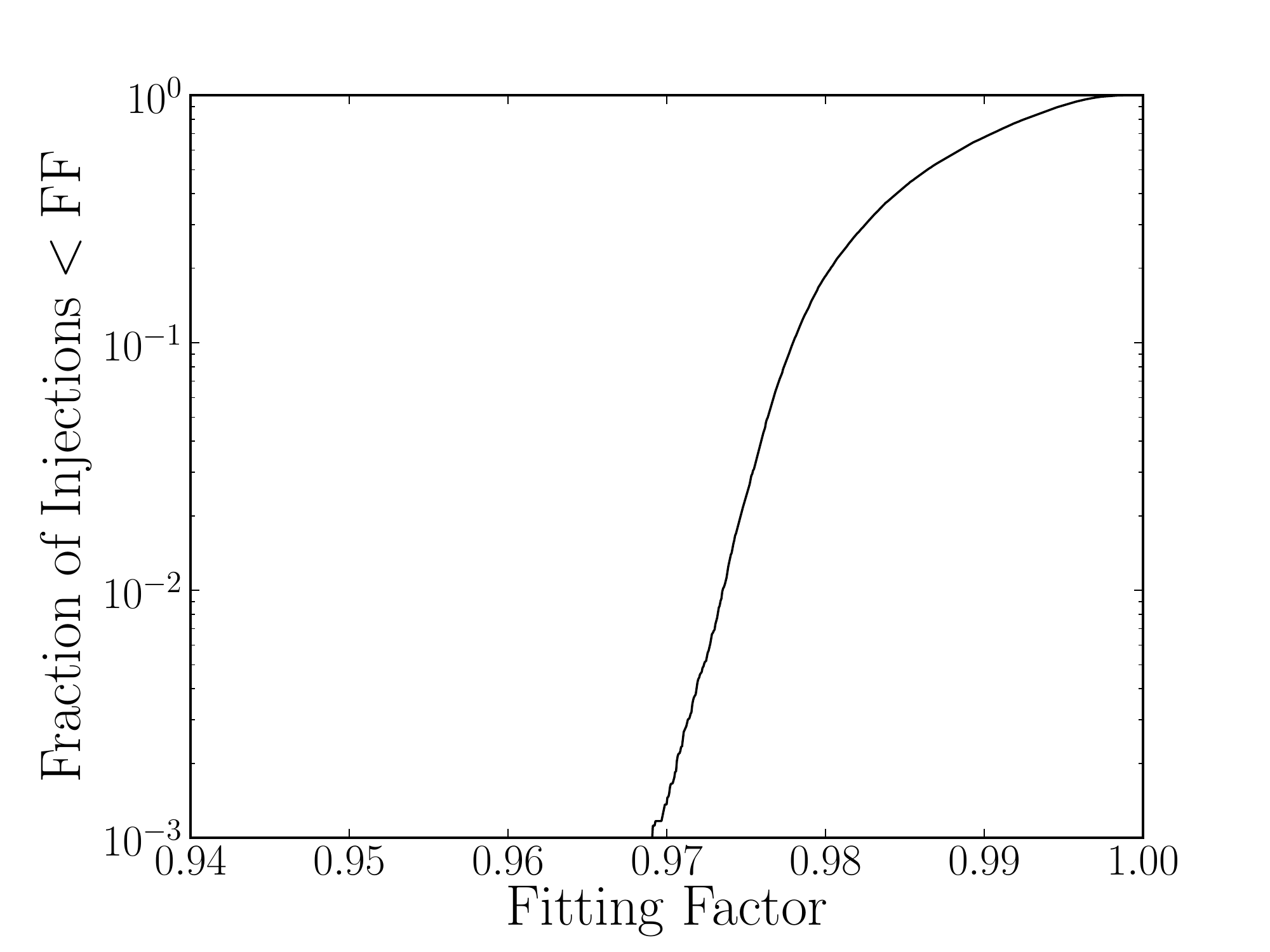}
\caption{\label{fig:anstar-aligned} The distribution of fitting factors obtained by searching
for aligned spin, binary neutron star systems, with spin magnitudes restricted to 0.4
using the aligned-spin BNS template bank described in section \ref{sec:param_space}
and the aLIGO, zero-detuned, high-power PSD with a 15Hz lower frequency cutoff.}
\end{figure}

We can verify that the template bank algorithm is working correctly by repeating the simulation described
in section \ref{ssec:nonspin_performance}, but evaluating the fitting factor between our bank of aligned-spin
template waveforms and a set of signals that is restricted to having spins that are (anti-)aligned with the
orbital angular momentum. The results of this simulation are shown in figure \ref{fig:anstar-aligned} and
one can see that with our bank we do not observe fitting factors lower than 0.97 when searching for aligned
spin BNS systems.

\begin{figure}
\includegraphics[width=0.45\textwidth]{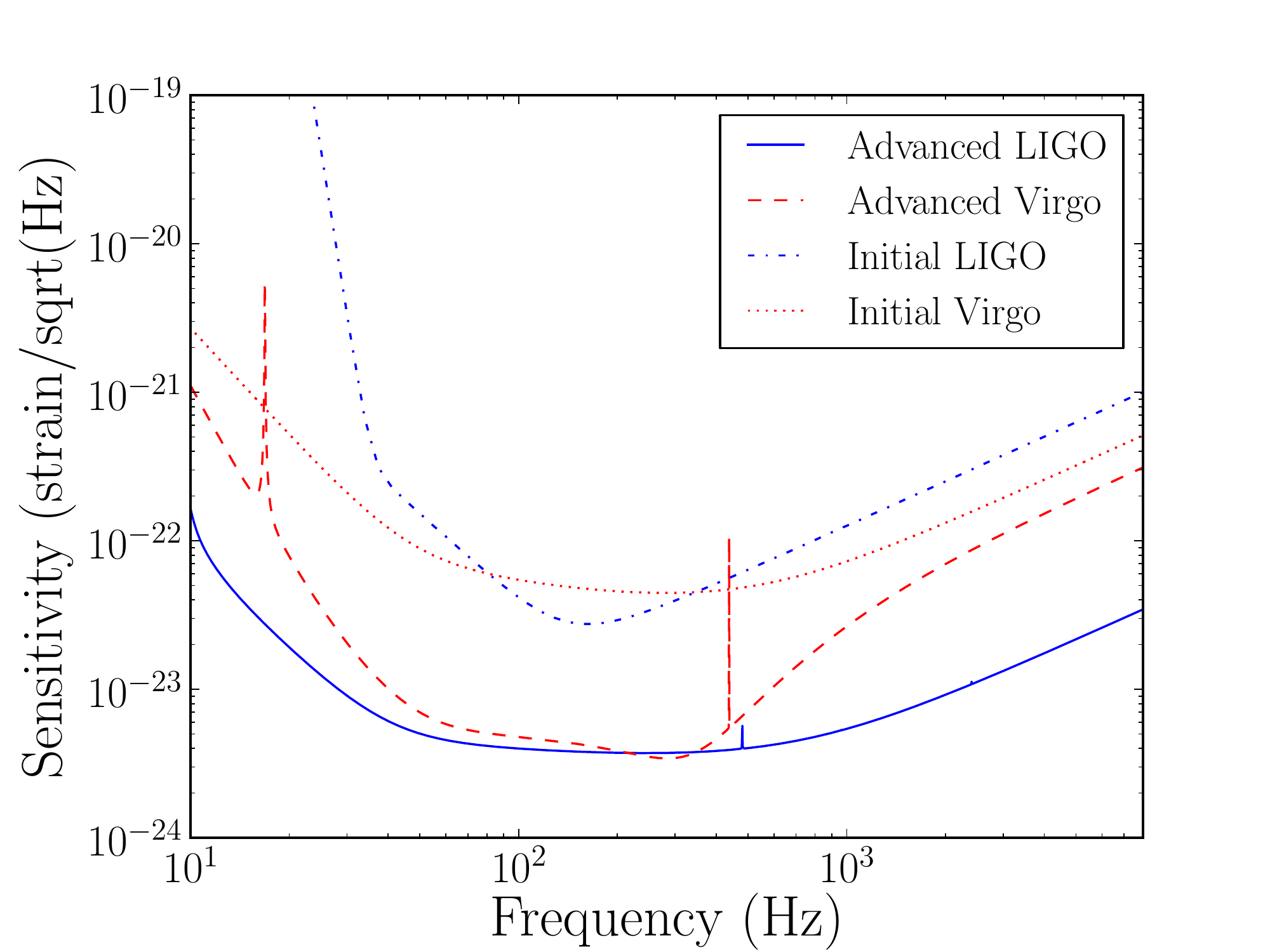}
\caption{\label{fig:asd_comparison} The amplitude spectral density for the aLIGO
zero-detuned high-power design sensitivity (blue solid curve), AdV design sensitivity
(red dashed curve), initial LIGO design sensitivity (blue bot-dash curve) and initial
Virgo design sensitivity (red dotted curve).}
\end{figure}

In the previous paragraphs we have restricted attention to the aLIGO
zero-detuned, high-power predicted sensitivity with a 15Hz lower frequency cut off. However,
we should verify that the conclusions we have drawn are valid for AdV, whose
PSD is different from that of aLIGO, as shown in Figure \ref{fig:asd_comparison}. Additionally we
should also show that the choice to use a 15Hz cut off in the aLIGO PSD does not affect
the conclusions made in this section. 

\begin{figure}
\includegraphics[width=0.45\textwidth]{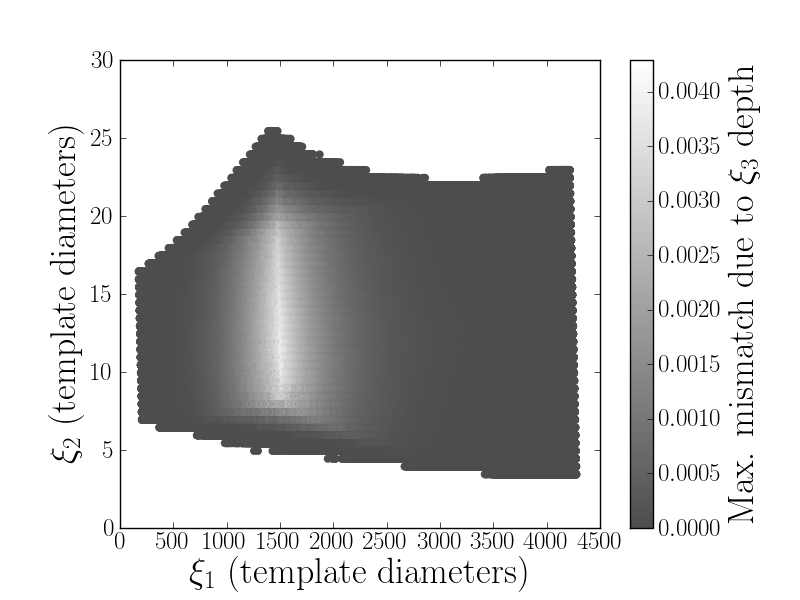}
\includegraphics[width=0.45\textwidth]{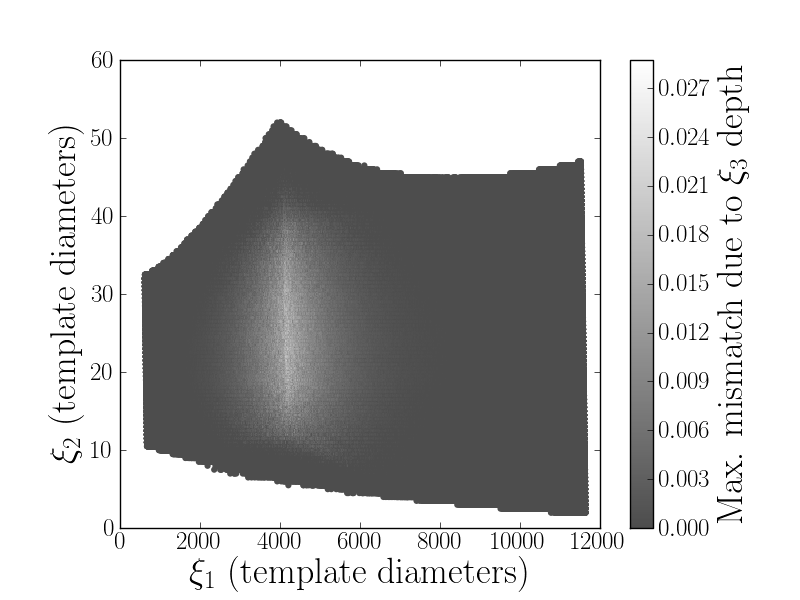}
\caption{\label{fig:param_space_mismatch_alt} The mismatch between the edge and centre of the
third dominant direction as a function of the first and second dominant
directions when using the Virgo noise curve (top) and when using the advanced LIGO noise curve
with a 10Hz lower frequency cut off (bottom). The $\xi_i$ coordinates have been scaled
such that one unit corresponds to the coverage diameter of a template
at 0.97 mismatch.
Plotted for a binary neutron star parameter space with spins restricted
to 0.4.}
\end{figure}

The process we described above is applicable for any PSD, and therefore we can use it directly
to determine the $\xi_i$ directions for the AdV PSD, or the aLIGO PSD with
a 10Hz lower frequency cutoff. In Figure \ref{fig:param_space_mismatch_alt} we plot
$\xi_1$ against $\xi_2$ for both PSDs while the color shows the mismatch between the center and edges in
the $\xi_3$ direction. This plot can be directly compared to Figure \ref{fig:param_space_mismatch}. We notice
that the size of the parameter space for the AdV PSD is significantly smaller than for the
aLIGO PSD in all 3 of the dominant directions. Therefore our conclusions for aLIGO are still
valid for AdV. Using our method we find that we require approximately 120,000 templates to cover the
parameter space for AdV, in comparison to approximately 520,000 templates for aLIGO. 

By comparing the results when using the aLIGO PSD with a 10Hz and 15Hz lower cut off we observe
that using a 10Hz lower frequency cut off will increase the number of necessary templates from $\sim520000$
to $\sim860000$. However the
shape of the parameter space, and thus our final conclusions, are unaffected when using a 10Hz lower
frequency cutoff. However, in this case we see larger mismatches due to the depth of $\xi_3$ and
therefore the process of stacking templates is important when using a 10Hz lower cut off. However, even
in this case, we do
not feel that the depth is large enough everywhere in the space to justify using a fully 3-dimensional placement
algorithm.

\begin{figure}
\includegraphics[width=0.45\textwidth]{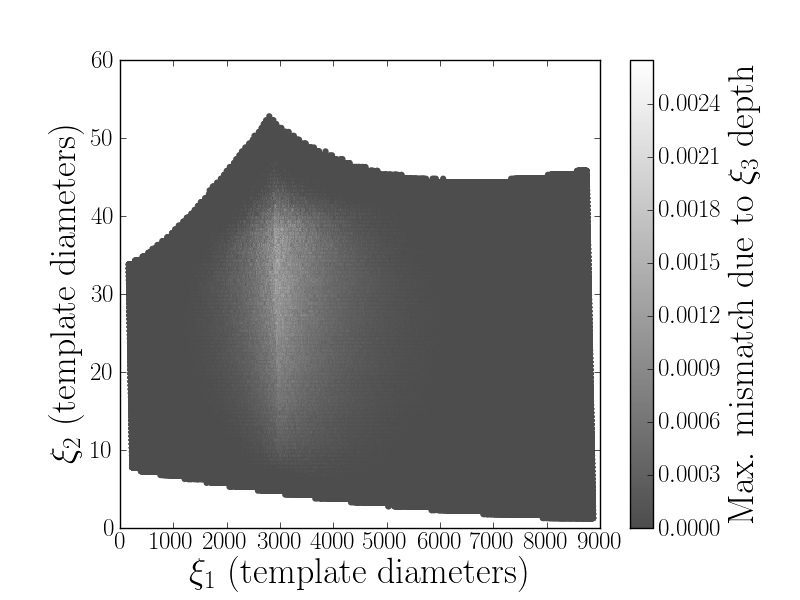}
\caption{\label{fig:higher_order_spin} The mismatch between the edge and centre of the
third dominant direction as a function of the first and second dominant
directions using waveforms incorporating the sub-dominant spin corrections to the orbital phase.
The $\xi_i$ coordinates have been scaled
such that one unit corresponds to the coverage diameter of a template
at 0.97 mismatch.
Plotted for a binary neutron star parameter space with spins restricted
to 0.4
using the zero-detuned, high-power aLIGO sensitivity curve with a 15Hz lower frequency cut off.}
\end{figure}

Finally, we wish to investigate the effect that the higher order spin contributions to the orbital phase
have on our method. To do this we repeat the process described above, but include the spin(1)-spin(1) and
spin(2)-spin(2) contributions to the $\sigma$ term at 2PN order and also the 2.5PN spin-orbit term as given
in \cite{Arun:2008kb}. In Figure \ref{fig:higher_order_spin} we plot $\xi_1$ against $\xi_2$ when these
higher order spin terms are included, the color shows the mismatch between the center and edges
in the $\xi_3$ direction. This plot can be directly compared to Figure \ref{fig:param_space_mismatch}.
By comparing these plots we can see that including the higher order spin terms has caused the parameter space
to have a larger extent in the $\xi_2$ direction. However, the depth of the space in the $\xi_3$ direction
has reduced by almost an order of magnitude. In this case the stacking process is not required and the resulting
bank consists of $\sim560000$ templates. 

\section{Comparison to alternative placement methods}

An alternative approach to template placement for aligned spin systems is to use templates
with ``unphysical'' values of the symmetric mass ratio, $\eta$.
That is, to use non-spinning templates, with the desired range of chirp
mass but where the range of $\eta$ values is extended to include both values of $\eta$ that are much lower than
the relevant parameter space and values of $\eta$ that are much higher,
including templates with $\eta$ greater than the physically possible limit of 0.25.

\begin{figure}
\includegraphics[width=0.45\textwidth]{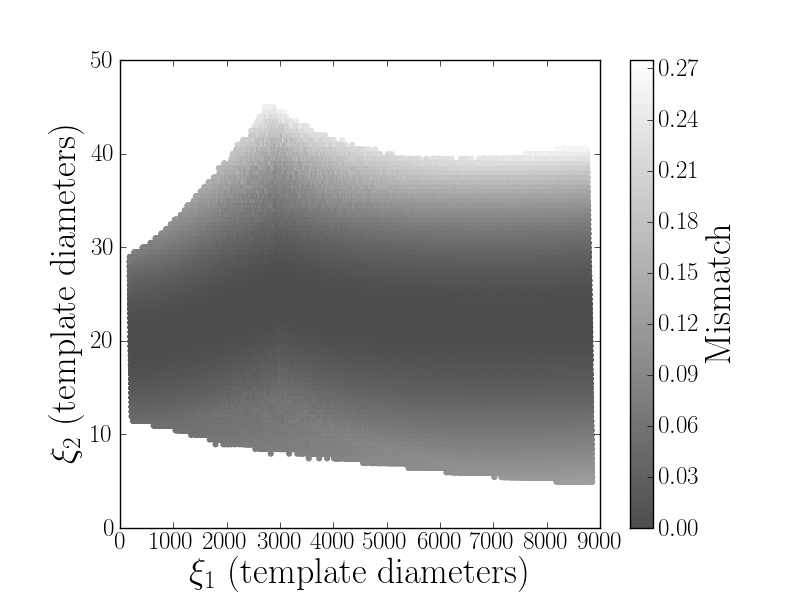} 
\caption{\label{fig:unphys_eta} The mismatch between unphysical $\eta$ and
aligned spin BNS templates as a function of the first and second dominant
directions. In making this plot we assume that $\xi_3$
has no depth in the aligned spin case by taking the central value where $\xi_3$ has a range of values.
This plot was generated
with spins restricted to 0.4 using the zero-detuned, high-power advanced LIGO
sensitivity curve with a 15Hz lower frequency cut off. 
}
\end{figure}

We can understand this unphysical $\eta$ approach in terms of our $\xi_i$ coordinate system by noting that
it is always possible to produce a template with any possible value
of $\xi_1$ and $\xi_2$ that is within the BNS parameter space, by using non-spinning templates
with unrestricted values of $\eta$.
By generating a set of templates in the $\xi_1$, $\xi_2$ directions,
where we restrict the chirp mass to be that possible for BNS systems, but where $\eta$
ranges from 0.1 to 0.7 we are able to cover the full physically possible space in $\xi_1$, $\xi_2$. However,
the disadvantage to using unphysical $\eta$ templates is that the points will not take the correct values
of $\xi_3$. The colorbar on Figure \ref{fig:unphys_eta} indicates the mismatch between unphysical $\eta$ templates and
aligned-spin templates as a function of $\xi_1$ and $\xi_2$. In making this plot we assume that $\xi_3$
has no depth in the aligned spin case by taking the central value where $\xi_3$ has a range of values.

While unphysical $\eta$ templates will produce an increase in efficiency when compared with non-spinning templates, the
method is not as efficient as the aligned spin geometrical placement we have described. In addition, both methods
require the same number of templates to cover the parameter space. Therefore, we would recommend using aligned spin templates
placed using our metric algorithm as opposed to unphysical $\eta$ templates.

Finally, we wish to compare the performance of this geometrical algorithm with the stochastic bank proposed in
\cite{Harry:2009ea,Babak:2008rb}. The stochastic placement works by randomly placing points within the parameter
space and rejecting points that are too ``close'' to points already in the bank. This
has the advantage that it is valid for any parameter space metric, so we could use any of the metrics discussed
above. However, it is more computationally efficient to use the Cartesian $\xi_i$ or $\mu_i$ coordinate system
rather than the non-Cartesian metric given above.

The disadvantage to a stochastic bank, when compared to a geometrically placed bank, is that it will require more
templates to achieve the same level of coverage \cite{Harry:2009ea,Manca:2009xw}.
For our parameter space, consisting of BNS signals with
component spins up to 0.4 and using the advanced LIGO zero-detuned high-power design curve with a 15Hz lower
frequency cut-off,
we found that the stochastic placement
produced a bank containing $\sim 750000$ templates, which is 44\% more than with the geometrical placement.
However, stochastic placement can still be used to place templates when no analytical metric is known, such
as when the merger becomes important. In such regions of parameter space, the stochastic placement may still be the best
algorithm to use to place a template bank.

\section{Performance of the aligned spin template bank}
\label{sec:aligned_spin_performance}

In this section we would like to investigate the improvement in the 
detection of generic BNS systems that results from using a template bank
that includes the dominant, non-precessing, spin effects. To do this we use the aligned spinning bank that
we detail in section \ref{sec:param_space} and compare this to the results of using a nonspinning bank 
as shown in section \ref{sec:spin_import}. 

\begin{figure}
\includegraphics[width=0.45\textwidth]{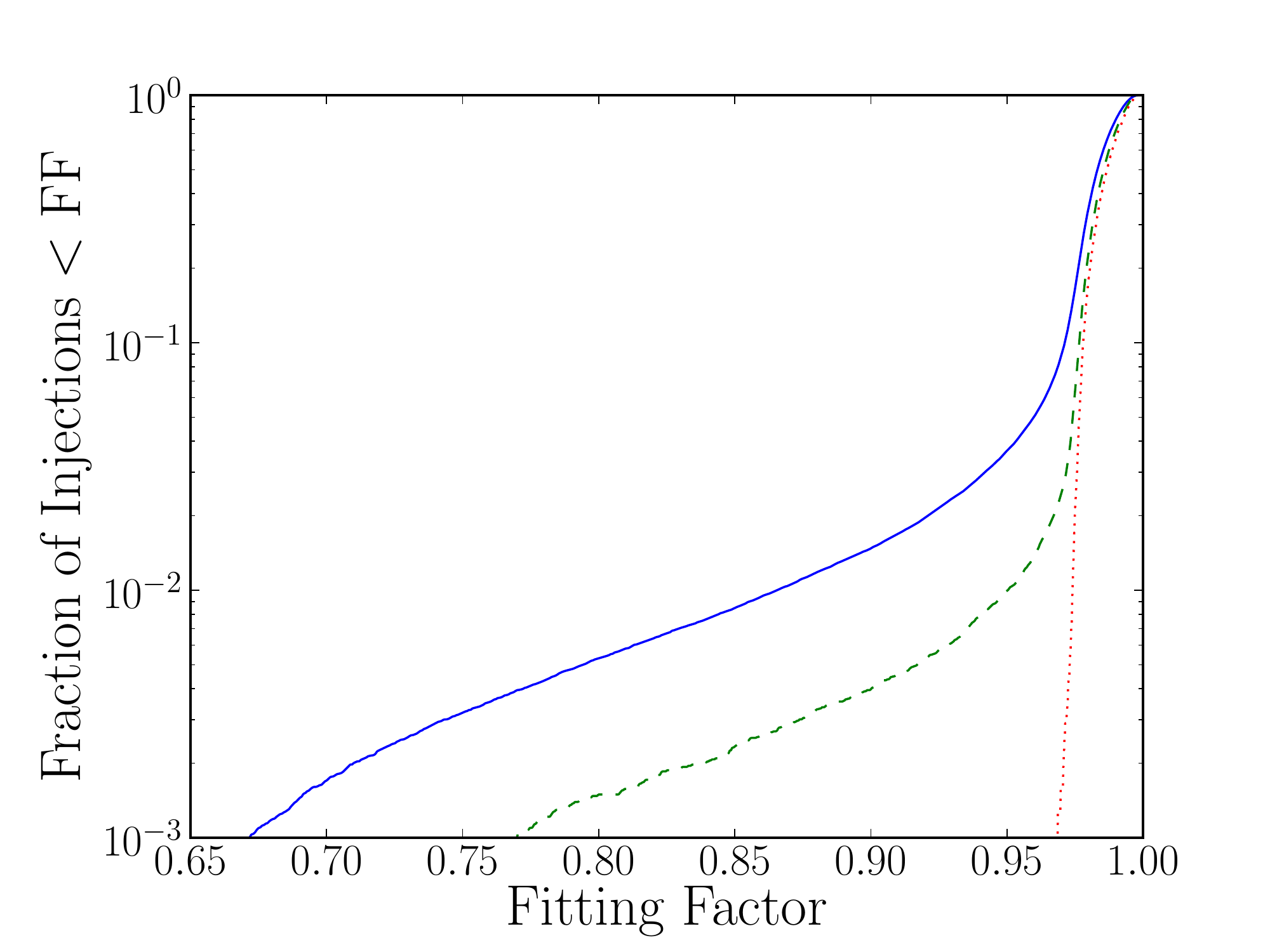}
\caption{\label{fig:anstar-prec} The distribution of fitting factors obtained by searching
for the precessing signals described in section \ref{ssec:nonspin_performance}
with component spins up to 0.4 (blue solid line), 0.2 (green dashed line), and 0.05 (red dotted line) using the aligned spin
BNS template bank described in section \ref{sec:param_space} and the advanced LIGO, zero-detuned,
high-power PSD with a 15Hz lower frequency cutoff.}
\end{figure}

Using our aligned spin template bank, we repeat the investigation from section \ref{sec:spin_import}. We create a 
population of source BNS signals identical to those used in \ref{ssec:nonspin_performance}, and compute the fitting factor
between these signals and the aligned spin template bank. The results of this are shown in FIG.\ref{fig:anstar-prec}.
To decrease the computational cost of this test, we only calculated the overlaps between a signal and templates that
were within a range of $\pm0.1M_{\odot}$ in chirp mass. This is reasonable because the overlap will decrease rapidly
with small changes in chirp mass,
therefore we expect templates with very different values of chirp mass to have low overlaps with each other. We verified
that this approach did not cause us to underestimate the fitting-factor of our banks.

\begin{figure}
\includegraphics[width=0.45\textwidth]{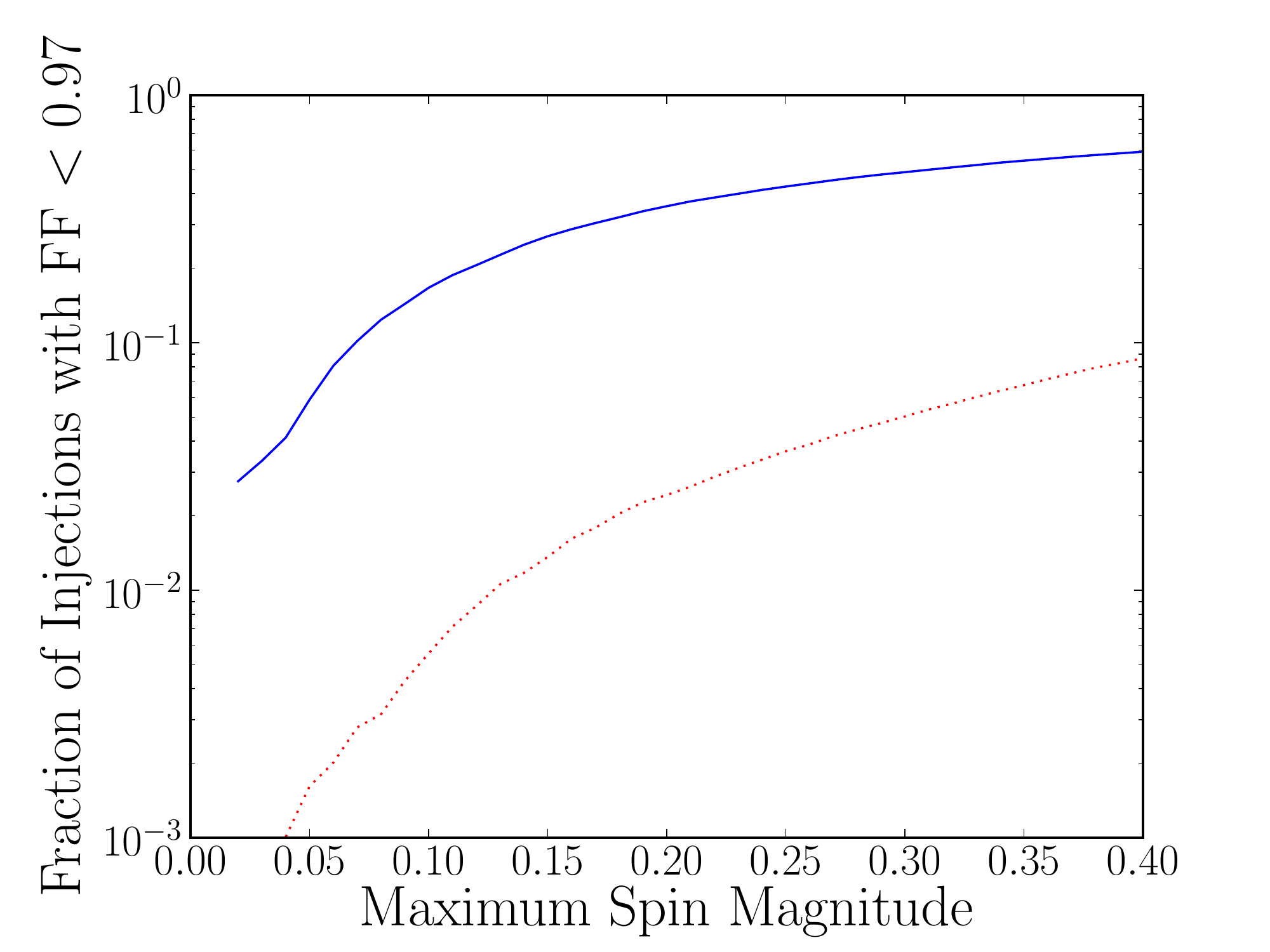}
\caption{\label{fig:anstar-st-spin} The fraction of the precessing signals described in
section \ref{ssec:nonspin_performance} recovered with a fitting factor less than 0.97 as
a function of the maximum component spin. Shown for the non-spinning
BNS template bank described in section \ref{ssec:nonspin_performance} (blue solid line),
and the aligned spin
BNS template bank described in section \ref{sec:param_space} (red dotted line). The advanced LIGO, zero-detuned,
high-power PSD with a 15Hz lower frequency cutoff was used when computing the fitting factors.}
\end{figure}

We can now compare the results obtained in this section, using our aligned-spin template bank, with the results obtained in section
\ref{sec:spin_import}, using a non-spinning template bank. One can clearly see an 
improvement in the distribution of fitting factors when using the aligned spin template bank. The fraction
of signals that fall below a fitting factor of 0.97, when the spin magnitudes are restricted to 0.4, falls from 59\% to 9\%.
We also see an  improvement for signals that have spin magnitudes restricted to 0.05, where the fraction of signals falling below a
fitting factor of 0.97 drops from 6\% to 0.2\%. We can also compare the performance of the aligned-spin bank to that of the
non-spinning bank as a function of the maximum spin magnitude,
as shown in Figure \ref{fig:anstar-st-spin}. From this Figure we can see that regardless of the maximum component
spin, the aligned spin bank will greatly reduce the number of signals recovered with fitting factors less than 0.97.


A small fraction of signals fall below a FF of 0.97, even when using the new aligned-spin template bank.
We expect that these poor matches with the aligned template bank are
due to precession. In general, precessional effects will not be important in BNS systems
as the orbital angular momentum is significantly larger than the component spins.
In such cases there is only a small angle between the total and orbital angular momenta
and precession has only a small effect on the waveform.

However, there is a small region of parameter space where precessional effects \textit{will}
have an effect for BNS systems.
Using the model of Ref.~\cite{Brown:2012gs}, applied to the small precession angles in BNS systems, 
we can predict for which systems precession will be most important.
The orientation of a precessing binary must be defined using the total angular momentum rather than the 
orbital angular momentum as done with non-precessing binaries. 
The orientations with the worst matches should be those where the system is edge-on 
(angular momentum perpendicular to the viewing direction) and where the detector is nearly insensitive 
to the plus polarization and only sees the cross polarization (a binary overhead of the detector would have 
its angular momentum oriented $45^{\circ}$ between the arms of the detector).
We find that
this is indeed the case; in fact, all cases with fitting factors less than 0.95 are close to
this configuration. All of these cases also have biases in the recovered mass and
spin parameters due to the secular effects of precession on the phasing of the waveform.

\section{Conclusion}
\label{sec:conclusion}

In this work we have investigated the effects of neglecting spin when
searching for binary neutron star systems in aLIGO and AdV. We have found
that, if component spins in binary neutron star systems are as large as 0.4,
then neutron star spin cannot be neglected, and there is a non-trivial loss in
signal-to-noise ratio even if the maximum spin is restricted to be less than
0.05.  We have developed a new algorithm for placing an aligned spin template
bank in the BNS parameter space.  We have shown that this bank works for
aligned spin systems and have demonstrated that it does significantly better
for generic, precessing BNS systems than the traditional non-spinning bank.
However, for the BNS aligned spin $\chi_i < 0.4$ parameter space the aligned
spin bank requires approximately five times as many templates as the
non-spinning bank. This increased number of templates will increase the
computational cost of the search and increase the number of background events,
so needs to be balanced against the potential gain in being able to cover a
larger region of parameter space. A further advantage of our method is the ease
with which it can be incorporated into existing or future search
pipelines, which include the use of signal-based vetoes~\cite{Allen:2004gu}
and coincidence algorithms~\cite{Robinson:2008}. In future work we will
investigate how this template bank performs in data from the aLIGO and AdV
detectors which includes non-Gaussian and non-stationary noise features.
Finally we note that the method proposed in this work should be applicable
wherever the TaylorF2 waveforms closely represent actual gravitational
waveforms. In a future work we will investigate how well this method performs
in the binary black hole and neutron-star, black-hole regions of the parameter space.
Wherever the TaylorF2 approximation begins to break down, a stochastic
bank placement may still be the most viable option.

\section*{Acknowledgements}

The authors are greatful to Stefan Ballmer, Stephen Fairhurst, Eliu Huerta, Drew Keppel, Prayush Kumar, Frank Ohme, Ben Owen,
Reinhard Prix, Peter
Saulson, B.S. Sathyaprakash, John Veitch, Matthew West and Karl Wette for helpful discussions. DB, IH and AN are
supported by NSF award PHY-0847611. IH and AN are also also supported by NSF
award PHY-0854812. AL was supported by NSF grant PHY-0855589 and the Max Planck Gesellschaft. DB and IH are also supported by a Cottrell Scholar award
from the Research Corporation for Science Advancement.  Computations used in
this work were performed on the Syracuse University Gravitation and Relativity
cluster, which is supported by NSF awards PHY-1040231, PHY-0600953 and
PHY-1104371.

\bibliography{references}

\begin{thebibliography}{56}%
\makeatletter
\providecommand \@ifxundefined [1]{%
 \@ifx{#1\undefined}
}%
\providecommand \@ifnum [1]{%
 \ifnum #1\expandafter \@firstoftwo
 \else \expandafter \@secondoftwo
 \fi
}%
\providecommand \@ifx [1]{%
 \ifx #1\expandafter \@firstoftwo
 \else \expandafter \@secondoftwo
 \fi
}%
\providecommand \natexlab [1]{#1}%
\providecommand \enquote  [1]{``#1''}%
\providecommand \bibnamefont  [1]{#1}%
\providecommand \bibfnamefont [1]{#1}%
\providecommand \citenamefont [1]{#1}%
\providecommand \href@noop [0]{\@secondoftwo}%
\providecommand \href [0]{\begingroup \@sanitize@url \@href}%
\providecommand \@href[1]{\@@startlink{#1}\@@href}%
\providecommand \@@href[1]{\endgroup#1\@@endlink}%
\providecommand \@sanitize@url [0]{\catcode `\\12\catcode `\$12\catcode
  `\&12\catcode `\#12\catcode `\^12\catcode `\_12\catcode `\%12\relax}%
\providecommand \@@startlink[1]{}%
\providecommand \@@endlink[0]{}%
\providecommand \url  [0]{\begingroup\@sanitize@url \@url }%
\providecommand \@url [1]{\endgroup\@href {#1}{\urlprefix }}%
\providecommand \urlprefix  [0]{URL }%
\providecommand \Eprint [0]{\href }%
\providecommand \doibase [0]{http://dx.doi.org/}%
\providecommand \selectlanguage [0]{\@gobble}%
\providecommand \bibinfo  [0]{\@secondoftwo}%
\providecommand \bibfield  [0]{\@secondoftwo}%
\providecommand \translation [1]{[#1]}%
\providecommand \BibitemOpen [0]{}%
\providecommand \bibitemStop [0]{}%
\providecommand \bibitemNoStop [0]{.\EOS\space}%
\providecommand \EOS [0]{\spacefactor3000\relax}%
\providecommand \BibitemShut  [1]{\csname bibitem#1\endcsname}%
\let\auto@bib@innerbib\@empty
\bibitem [{\citenamefont {Harry}\ \emph {et~al.}(2010)\citenamefont {Harry}
  \emph {et~al.}}]{Harry:2010zz}%
  \BibitemOpen
  \bibfield  {author} {\bibinfo {author} {\bibfnamefont {G.~M.}\ \bibnamefont
  {Harry}} \emph {et~al.},\ }\href {\doibase 10.1088/0264-9381/27/8/084006}
  {\bibfield  {journal} {\bibinfo  {journal} {Class. Quant. Grav.}\ }\textbf
  {\bibinfo {volume} {27}},\ \bibinfo {pages} {084006} (\bibinfo {year}
  {2010})}\BibitemShut {NoStop}%
\bibitem [{\citenamefont {Acernese}\ \emph {et~al.}(2009)\citenamefont
  {Acernese} \emph {et~al.}}]{aVirgo}%
  \BibitemOpen
  \bibfield  {author} {\bibinfo {author} {\bibfnamefont {F.}~\bibnamefont
  {Acernese}} \emph {et~al.},\ }\href@noop {} {\  (\bibinfo {year} {2009})},\
  \bibinfo {note} {{Virgo Technical Report 0027A-09}}\BibitemShut {NoStop}%
\bibitem [{\citenamefont {Abadie}\ \emph
  {et~al.}(2010{\natexlab{a}})\citenamefont {Abadie} \emph
  {et~al.}}]{Abadie:2010cf}%
  \BibitemOpen
  \bibfield  {author} {\bibinfo {author} {\bibfnamefont {J.}~\bibnamefont
  {Abadie}} \emph {et~al.},\ }\href {\doibase 10.1088/0264-9381/27/17/173001}
  {\bibfield  {journal} {\bibinfo  {journal} {Class.Quant.Grav.}\ }\textbf
  {\bibinfo {volume} {27}},\ \bibinfo {pages} {173001} (\bibinfo {year}
  {2010}{\natexlab{a}})}\BibitemShut {NoStop}%
\bibitem [{\citenamefont {Blanchet}(2006)}]{Blanchet:2006zz}%
  \BibitemOpen
  \bibfield  {author} {\bibinfo {author} {\bibfnamefont {L.}~\bibnamefont
  {Blanchet}},\ }\href@noop {} {\bibfield  {journal} {\bibinfo  {journal}
  {Living Rev.Rel.}\ }\textbf {\bibinfo {volume} {9}} (\bibinfo {year}
  {2006})}\BibitemShut {NoStop}%
\bibitem [{\citenamefont {Peters}\ and\ \citenamefont
  {Mathews}(1963)}]{Peters:1963ux}%
  \BibitemOpen
  \bibfield  {author} {\bibinfo {author} {\bibfnamefont {P.~C.}\ \bibnamefont
  {Peters}}\ and\ \bibinfo {author} {\bibfnamefont {J.}~\bibnamefont
  {Mathews}},\ }\href {\doibase 10.1103/PhysRev.131.435} {\bibfield  {journal}
  {\bibinfo  {journal} {Phys. Rev.}\ }\textbf {\bibinfo {volume} {131}},\
  \bibinfo {pages} {435} (\bibinfo {year} {1963})}\BibitemShut {NoStop}%
\bibitem [{\citenamefont {Blanchet}\ \emph
  {et~al.}(1995{\natexlab{a}})\citenamefont {Blanchet}, \citenamefont
  {Damour},\ and\ \citenamefont {Iyer}}]{Blanchet:1995fg}%
  \BibitemOpen
  \bibfield  {author} {\bibinfo {author} {\bibfnamefont {L.}~\bibnamefont
  {Blanchet}}, \bibinfo {author} {\bibfnamefont {T.}~\bibnamefont {Damour}}, \
  and\ \bibinfo {author} {\bibfnamefont {B.~R.}\ \bibnamefont {Iyer}},\
  }\href@noop {} {\bibfield  {journal} {\bibinfo  {journal} {Phys. Rev.}\
  }\textbf {\bibinfo {volume} {D51}},\ \bibinfo {pages} {5360} (\bibinfo {year}
  {1995}{\natexlab{a}})}\BibitemShut {NoStop}%
\bibitem [{\citenamefont {Blanchet}\ \emph
  {et~al.}(1995{\natexlab{b}})\citenamefont {Blanchet}, \citenamefont {Damour},
  \citenamefont {Iyer}, \citenamefont {Will},\ and\ \citenamefont
  {Wiseman}}]{Blanchet:1995ez}%
  \BibitemOpen
  \bibfield  {author} {\bibinfo {author} {\bibfnamefont {L.}~\bibnamefont
  {Blanchet}}, \bibinfo {author} {\bibfnamefont {T.}~\bibnamefont {Damour}},
  \bibinfo {author} {\bibfnamefont {B.~R.}\ \bibnamefont {Iyer}}, \bibinfo
  {author} {\bibfnamefont {C.~M.}\ \bibnamefont {Will}}, \ and\ \bibinfo
  {author} {\bibfnamefont {A.~G.}\ \bibnamefont {Wiseman}},\ }\href@noop {}
  {\bibfield  {journal} {\bibinfo  {journal} {Phys. Rev. Lett.}\ }\textbf
  {\bibinfo {volume} {74}},\ \bibinfo {pages} {3515} (\bibinfo {year}
  {1995}{\natexlab{b}})}\BibitemShut {NoStop}%
\bibitem [{\citenamefont {Blanchet}\ \emph {et~al.}(1996)\citenamefont
  {Blanchet}, \citenamefont {Iyer}, \citenamefont {Will},\ and\ \citenamefont
  {Wiseman}}]{BIWW96}%
  \BibitemOpen
  \bibfield  {author} {\bibinfo {author} {\bibfnamefont {L.}~\bibnamefont
  {Blanchet}}, \bibinfo {author} {\bibfnamefont {B.~R.}\ \bibnamefont {Iyer}},
  \bibinfo {author} {\bibfnamefont {C.~M.}\ \bibnamefont {Will}}, \ and\
  \bibinfo {author} {\bibfnamefont {A.~G.}\ \bibnamefont {Wiseman}},\
  }\href@noop {} {\bibfield  {journal} {\bibinfo  {journal} {Class. Quantum
  Grav.}\ }\textbf {\bibinfo {volume} {13}},\ \bibinfo {pages} {575} (\bibinfo
  {year} {1996})}\BibitemShut {NoStop}%
\bibitem [{\citenamefont {Wiseman}(1993)}]{Wi93}%
  \BibitemOpen
  \bibfield  {author} {\bibinfo {author} {\bibfnamefont {A.~G.}\ \bibnamefont
  {Wiseman}},\ }\href@noop {} {\bibfield  {journal} {\bibinfo  {journal} {Phys.
  Rev. D}\ }\textbf {\bibinfo {volume} {48}},\ \bibinfo {pages} {4757}
  (\bibinfo {year} {1993})}\BibitemShut {NoStop}%
\bibitem [{\citenamefont {Blanchet}\ \emph {et~al.}(2002)\citenamefont
  {Blanchet}, \citenamefont {Faye}, \citenamefont {Iyer},\ and\ \citenamefont
  {Joguet}}]{BFIJ02}%
  \BibitemOpen
  \bibfield  {author} {\bibinfo {author} {\bibfnamefont {L.}~\bibnamefont
  {Blanchet}}, \bibinfo {author} {\bibfnamefont {G.}~\bibnamefont {Faye}},
  \bibinfo {author} {\bibfnamefont {B.~R.}\ \bibnamefont {Iyer}}, \ and\
  \bibinfo {author} {\bibfnamefont {B.}~\bibnamefont {Joguet}},\ }\href@noop {}
  {\bibfield  {journal} {\bibinfo  {journal} {Phys. Rev. D}\ }\textbf {\bibinfo
  {volume} {65}},\ \bibinfo {pages} {061501(R)} (\bibinfo {year} {2002})},\
  \bibinfo {note} {{Erratum-ibid~{\bf 71}, 129902(E) (2005)}}\BibitemShut
  {NoStop}%
\bibitem [{\citenamefont {Blanchet}\ \emph {et~al.}(2004)\citenamefont
  {Blanchet}, \citenamefont {Damour}, \citenamefont {Esposito-Farese},\ and\
  \citenamefont {Iyer}}]{Blanchet:2004ek}%
  \BibitemOpen
  \bibfield  {author} {\bibinfo {author} {\bibfnamefont {L.}~\bibnamefont
  {Blanchet}}, \bibinfo {author} {\bibfnamefont {T.}~\bibnamefont {Damour}},
  \bibinfo {author} {\bibfnamefont {G.}~\bibnamefont {Esposito-Farese}}, \ and\
  \bibinfo {author} {\bibfnamefont {B.~R.}\ \bibnamefont {Iyer}},\ }\href@noop
  {} {\bibfield  {journal} {\bibinfo  {journal} {Phys. Rev. Lett.}\ }\textbf
  {\bibinfo {volume} {93}},\ \bibinfo {pages} {091101} (\bibinfo {year}
  {2004})}\BibitemShut {NoStop}%
\bibitem [{\citenamefont {Kidder}\ \emph {et~al.}(1993)\citenamefont {Kidder},
  \citenamefont {Will},\ and\ \citenamefont {Wiseman}}]{Kidder:1992fr}%
  \BibitemOpen
  \bibfield  {author} {\bibinfo {author} {\bibfnamefont {L.~E.}\ \bibnamefont
  {Kidder}}, \bibinfo {author} {\bibfnamefont {C.~M.}\ \bibnamefont {Will}}, \
  and\ \bibinfo {author} {\bibfnamefont {A.~G.}\ \bibnamefont {Wiseman}},\
  }\href@noop {} {\bibfield  {journal} {\bibinfo  {journal} {Phys. Rev.}\
  }\textbf {\bibinfo {volume} {D47}},\ \bibinfo {pages} {4183} (\bibinfo {year}
  {1993})}\BibitemShut {NoStop}%
\bibitem [{\citenamefont {Apostolatos}\ \emph {et~al.}(1994)\citenamefont
  {Apostolatos}, \citenamefont {Cutler}, \citenamefont {Sussman},\ and\
  \citenamefont {Thorne}}]{Apostolatos:1994mx}%
  \BibitemOpen
  \bibfield  {author} {\bibinfo {author} {\bibfnamefont {T.~A.}\ \bibnamefont
  {Apostolatos}}, \bibinfo {author} {\bibfnamefont {C.}~\bibnamefont {Cutler}},
  \bibinfo {author} {\bibfnamefont {G.~J.}\ \bibnamefont {Sussman}}, \ and\
  \bibinfo {author} {\bibfnamefont {K.~S.}\ \bibnamefont {Thorne}},\ }\href
  {\doibase 10.1103/PhysRevD.49.6274} {\bibfield  {journal} {\bibinfo
  {journal} {Phys. Rev.}\ }\textbf {\bibinfo {volume} {D49}},\ \bibinfo {pages}
  {6274} (\bibinfo {year} {1994})}\BibitemShut {NoStop}%
\bibitem [{\citenamefont {Kidder}(1995)}]{Kidder:1995zr}%
  \BibitemOpen
  \bibfield  {author} {\bibinfo {author} {\bibfnamefont {L.~E.}\ \bibnamefont
  {Kidder}},\ }\href@noop {} {\bibfield  {journal} {\bibinfo  {journal} {Phys.
  Rev.}\ }\textbf {\bibinfo {volume} {D52}},\ \bibinfo {pages} {821} (\bibinfo
  {year} {1995})}\BibitemShut {NoStop}%
\bibitem [{\citenamefont {Blanchet}\ \emph {et~al.}(2006)\citenamefont
  {Blanchet}, \citenamefont {Buonanno},\ and\ \citenamefont
  {Faye}}]{Blanchet:2006gy}%
  \BibitemOpen
  \bibfield  {author} {\bibinfo {author} {\bibfnamefont {L.}~\bibnamefont
  {Blanchet}}, \bibinfo {author} {\bibfnamefont {A.}~\bibnamefont {Buonanno}},
  \ and\ \bibinfo {author} {\bibfnamefont {G.}~\bibnamefont {Faye}},\
  }\href@noop {} {\bibfield  {journal} {\bibinfo  {journal} {Phys.Rev.}\
  }\textbf {\bibinfo {volume} {D74}},\ \bibinfo {pages} {104034} (\bibinfo
  {year} {2006})}\BibitemShut {NoStop}%
\bibitem [{\citenamefont {Lo}\ and\ \citenamefont {Lin}(2011)}]{Lo:2010bj}%
  \BibitemOpen
  \bibfield  {author} {\bibinfo {author} {\bibfnamefont {K.-W.}\ \bibnamefont
  {Lo}}\ and\ \bibinfo {author} {\bibfnamefont {L.-M.}\ \bibnamefont {Lin}},\
  }\href {\doibase 10.1088/0004-637X/728/1/12} {\bibfield  {journal} {\bibinfo
  {journal} {Astrophys.J.}\ }\textbf {\bibinfo {volume} {728}},\ \bibinfo
  {pages} {12} (\bibinfo {year} {2011})}\BibitemShut {NoStop}%
\bibitem [{\citenamefont {Lorimer}(2008)}]{Lorimer:2008se}%
  \BibitemOpen
  \bibfield  {author} {\bibinfo {author} {\bibfnamefont {D.}~\bibnamefont
  {Lorimer}},\ }\href@noop {} {\bibfield  {journal} {\bibinfo  {journal}
  {Living Rev.Rel.}\ }\textbf {\bibinfo {volume} {11}},\ \bibinfo {pages} {8}
  (\bibinfo {year} {2008})}\BibitemShut {NoStop}%
\bibitem [{\citenamefont {Mandel}\ and\ \citenamefont
  {O'Shaughnessy}(2010)}]{Mandel:2009nx}%
  \BibitemOpen
  \bibfield  {author} {\bibinfo {author} {\bibfnamefont {I.}~\bibnamefont
  {Mandel}}\ and\ \bibinfo {author} {\bibfnamefont {R.}~\bibnamefont
  {O'Shaughnessy}},\ }\href {\doibase 10.1088/0264-9381/27/11/114007}
  {\bibfield  {journal} {\bibinfo  {journal} {Class. Quant. Grav.}\ }\textbf
  {\bibinfo {volume} {27}},\ \bibinfo {pages} {114007} (\bibinfo {year}
  {2010})}\BibitemShut {NoStop}%
\bibitem [{\citenamefont {Bildsten}\ \emph {et~al.}(1997)\citenamefont
  {Bildsten} \emph {et~al.}}]{Bildsten:1997vw}%
  \BibitemOpen
  \bibfield  {author} {\bibinfo {author} {\bibfnamefont {L.}~\bibnamefont
  {Bildsten}} \emph {et~al.},\ }\href@noop {} {\bibfield  {journal} {\bibinfo
  {journal} {Astrophys J. Supp.}\ }\textbf {\bibinfo {volume} {113}},\ \bibinfo
  {pages} {367} (\bibinfo {year} {1997})}\BibitemShut {NoStop}%
\bibitem [{\citenamefont {{Chakrabarty}}(2008)}]{Chakrabarty:2008gz}%
  \BibitemOpen
  \bibfield  {author} {\bibinfo {author} {\bibfnamefont {D.}~\bibnamefont
  {{Chakrabarty}}},\ }\href@noop {} {\bibfield  {journal} {\bibinfo  {journal}
  {AIP Conference Series}\ }\bibinfo {series} {American Institute of Physics
  Conference Series},\ \textbf {\bibinfo {volume} {1068}},\ \bibinfo {pages}
  {67} (\bibinfo {year} {2008})}\BibitemShut {NoStop}%
\bibitem [{\citenamefont {Burgay}\ \emph {et~al.}(2003)\citenamefont {Burgay}
  \emph {et~al.}}]{Burgay:2003jj}%
  \BibitemOpen
  \bibfield  {author} {\bibinfo {author} {\bibfnamefont {M.}~\bibnamefont
  {Burgay}} \emph {et~al.},\ }\href {\doibase 10.1038/nature02124} {\bibfield
  {journal} {\bibinfo  {journal} {Nature}\ }\textbf {\bibinfo {volume} {426}},\
  \bibinfo {pages} {531} (\bibinfo {year} {2003})}\BibitemShut {NoStop}%
\bibitem [{\citenamefont {Grindlay}\ \emph {et~al.}(2006)\citenamefont
  {Grindlay}, \citenamefont {Zwart},\ and\ \citenamefont
  {McMillan}}]{Grindlay:2005ym}%
  \BibitemOpen
  \bibfield  {author} {\bibinfo {author} {\bibfnamefont {J.}~\bibnamefont
  {Grindlay}}, \bibinfo {author} {\bibfnamefont {S.~P.}\ \bibnamefont {Zwart}},
  \ and\ \bibinfo {author} {\bibfnamefont {S.}~\bibnamefont {McMillan}},\
  }\href@noop {} {\bibfield  {journal} {\bibinfo  {journal} {Nature Physics}\
  }\textbf {\bibinfo {volume} {2}},\ \bibinfo {pages} {116} (\bibinfo {year}
  {2006})}\BibitemShut {NoStop}%
\bibitem [{\citenamefont {Farr}\ \emph {et~al.}(2011)\citenamefont {Farr} \emph
  {et~al.}}]{Farr:2011gs}%
  \BibitemOpen
  \bibfield  {author} {\bibinfo {author} {\bibfnamefont {W.~M.}\ \bibnamefont
  {Farr}} \emph {et~al.},\ }\href@noop {} {\bibfield  {journal} {\bibinfo
  {journal} {Astrophys.J.}\ }\textbf {\bibinfo {volume} {742}},\ \bibinfo
  {pages} {81} (\bibinfo {year} {2011})}\BibitemShut {NoStop}%
\bibitem [{\citenamefont {Allen}\ \emph {et~al.}(2005)\citenamefont {Allen}
  \emph {et~al.}}]{Allen:2005fk}%
  \BibitemOpen
  \bibfield  {author} {\bibinfo {author} {\bibfnamefont {B.}~\bibnamefont
  {Allen}} \emph {et~al.},\ }\href@noop {} {\  (\bibinfo {year} {2005})},\
  \Eprint {http://arxiv.org/abs/gr-qc/0509116} {arXiv:gr-qc/0509116 [gr-qc]}
  \BibitemShut {NoStop}%
\bibitem [{\citenamefont {Owen}\ and\ \citenamefont
  {Sathyaprakash}(1999)}]{OwenSathyaprakash98}%
  \BibitemOpen
  \bibfield  {author} {\bibinfo {author} {\bibfnamefont {B.~J.}\ \bibnamefont
  {Owen}}\ and\ \bibinfo {author} {\bibfnamefont {B.~S.}\ \bibnamefont
  {Sathyaprakash}},\ }\href@noop {} {\bibfield  {journal} {\bibinfo  {journal}
  {Phys. Rev.}\ }\textbf {\bibinfo {volume} {D60}},\ \bibinfo {pages} {022002}
  (\bibinfo {year} {1999})}\BibitemShut {NoStop}%
\bibitem [{\citenamefont {Marion}\ \emph {et~al.}(2004)\citenamefont {Marion}
  \emph {et~al.}}]{Marion:2004}%
  \BibitemOpen
  \bibfield  {author} {\bibinfo {author} {\bibfnamefont {F.}~\bibnamefont
  {Marion}} \emph {et~al.} (\bibinfo {collaboration} {Virgo Collaboration}),\
  }in\ \href@noop {} {\emph {\bibinfo {booktitle} {Proceedings of the
  Rencontres de Moriond 2003}}}\ (\bibinfo {year} {2004})\BibitemShut {NoStop}%
\bibitem [{\citenamefont {Cannon}\ \emph {et~al.}(2010)\citenamefont {Cannon}
  \emph {et~al.}}]{Cannon:2010qh}%
  \BibitemOpen
  \bibfield  {author} {\bibinfo {author} {\bibfnamefont {K.}~\bibnamefont
  {Cannon}} \emph {et~al.},\ }\href@noop {} {\bibfield  {journal} {\bibinfo
  {journal} {Phys.Rev.}\ }\textbf {\bibinfo {volume} {D82}},\ \bibinfo {pages}
  {044025} (\bibinfo {year} {2010})}\BibitemShut {NoStop}%
\bibitem [{\citenamefont {Abadie}\ \emph {et~al.}(2012)\citenamefont {Abadie}
  \emph {et~al.}}]{Abadie:2011nz}%
  \BibitemOpen
  \bibfield  {author} {\bibinfo {author} {\bibfnamefont {J.}~\bibnamefont
  {Abadie}} \emph {et~al.},\ }\href@noop {} {\bibfield  {journal} {\bibinfo
  {journal} {Phys.Rev.}\ }\textbf {\bibinfo {volume} {D85}},\ \bibinfo {pages}
  {082002} (\bibinfo {year} {2012})}\BibitemShut {NoStop}%
\bibitem [{\citenamefont {Apostolatos}(1996)}]{Apostolatos:1996rf}%
  \BibitemOpen
  \bibfield  {author} {\bibinfo {author} {\bibfnamefont {T.~A.}\ \bibnamefont
  {Apostolatos}},\ }\href {\doibase 10.1103/PhysRevD.54.2421} {\bibfield
  {journal} {\bibinfo  {journal} {Phys.Rev.}\ }\textbf {\bibinfo {volume}
  {D54}},\ \bibinfo {pages} {2421} (\bibinfo {year} {1996})}\BibitemShut
  {NoStop}%
\bibitem [{\citenamefont {Ajith}(2011)}]{Ajith:2011ec}%
  \BibitemOpen
  \bibfield  {author} {\bibinfo {author} {\bibfnamefont {P.}~\bibnamefont
  {Ajith}},\ }\href {\doibase 10.1103/PhysRevD.84.084037} {\bibfield  {journal}
  {\bibinfo  {journal} {Phys.Rev.}\ }\textbf {\bibinfo {volume} {D84}},\
  \bibinfo {pages} {084037} (\bibinfo {year} {2011})}\BibitemShut {NoStop}%
\bibitem [{\citenamefont {Babak}\ \emph {et~al.}(2006)\citenamefont {Babak}
  \emph {et~al.}}]{Bank06}%
  \BibitemOpen
  \bibfield  {author} {\bibinfo {author} {\bibfnamefont {S.}~\bibnamefont
  {Babak}} \emph {et~al.},\ }\href@noop {} {\bibfield  {journal} {\bibinfo
  {journal} {Class. Quant. Grav.}\ }\textbf {\bibinfo {volume} {23}},\ \bibinfo
  {pages} {5477} (\bibinfo {year} {2006})}\BibitemShut {NoStop}%
\bibitem [{\citenamefont {Harry}\ \emph {et~al.}(2009)\citenamefont {Harry},
  \citenamefont {Allen},\ and\ \citenamefont {Sathyaprakash}}]{Harry:2009ea}%
  \BibitemOpen
  \bibfield  {author} {\bibinfo {author} {\bibfnamefont {I.~W.}\ \bibnamefont
  {Harry}}, \bibinfo {author} {\bibfnamefont {B.}~\bibnamefont {Allen}}, \ and\
  \bibinfo {author} {\bibfnamefont {B.~S.}\ \bibnamefont {Sathyaprakash}},\
  }\href@noop {} {\bibfield  {journal} {\bibinfo  {journal} {Phys. Rev.}\
  }\textbf {\bibinfo {volume} {D80}},\ \bibinfo {pages} {104014} (\bibinfo
  {year} {2009})}\BibitemShut {NoStop}%
\bibitem [{\citenamefont {Apostolatos}(1995)}]{Apostolatos:1995pj}%
  \BibitemOpen
  \bibfield  {author} {\bibinfo {author} {\bibfnamefont {T.~A.}\ \bibnamefont
  {Apostolatos}},\ }\href {\doibase 10.1103/PhysRevD.52.605} {\bibfield
  {journal} {\bibinfo  {journal} {Phys.Rev.}\ }\textbf {\bibinfo {volume}
  {D52}},\ \bibinfo {pages} {605} (\bibinfo {year} {1995})}\BibitemShut
  {NoStop}%
\bibitem [{\citenamefont {Cutler}\ and\ \citenamefont {Flanagan}(1994)}]{CF94}%
  \BibitemOpen
  \bibfield  {author} {\bibinfo {author} {\bibfnamefont {C.}~\bibnamefont
  {Cutler}}\ and\ \bibinfo {author} {\bibfnamefont {E.~E.}\ \bibnamefont
  {Flanagan}},\ }\href@noop {} {\bibfield  {journal} {\bibinfo  {journal}
  {Phys. Rev. D}\ }\textbf {\bibinfo {volume} {49}},\ \bibinfo {pages} {2658}
  (\bibinfo {year} {1994})}\BibitemShut {NoStop}%
\bibitem [{\citenamefont {Cutler}\ \emph {et~al.}(1993)\citenamefont {Cutler}
  \emph {et~al.}}]{Cutler:1992tc}%
  \BibitemOpen
  \bibfield  {author} {\bibinfo {author} {\bibfnamefont {C.}~\bibnamefont
  {Cutler}} \emph {et~al.},\ }\href@noop {} {\bibfield  {journal} {\bibinfo
  {journal} {Phys. Rev. Lett.}\ }\textbf {\bibinfo {volume} {70}},\ \bibinfo
  {pages} {2984} (\bibinfo {year} {1993})}\BibitemShut {NoStop}%
\bibitem [{\citenamefont {Droz}\ \emph {et~al.}(1999)\citenamefont {Droz},
  \citenamefont {Knapp}, \citenamefont {Poisson},\ and\ \citenamefont
  {Owen}}]{DPK99}%
  \BibitemOpen
  \bibfield  {author} {\bibinfo {author} {\bibfnamefont {S.}~\bibnamefont
  {Droz}}, \bibinfo {author} {\bibfnamefont {D.~J.}\ \bibnamefont {Knapp}},
  \bibinfo {author} {\bibfnamefont {E.}~\bibnamefont {Poisson}}, \ and\
  \bibinfo {author} {\bibfnamefont {B.~J.}\ \bibnamefont {Owen}},\ }\href@noop
  {} {\bibfield  {journal} {\bibinfo  {journal} {Phys. Rev. D}\ }\textbf
  {\bibinfo {volume} {59}},\ \bibinfo {pages} {124016} (\bibinfo {year}
  {1999})}\BibitemShut {NoStop}%
\bibitem [{\citenamefont {Cokelaer}(2007)}]{Cokelaer:2007kx}%
  \BibitemOpen
  \bibfield  {author} {\bibinfo {author} {\bibfnamefont {T.}~\bibnamefont
  {Cokelaer}},\ }\href@noop {} {\bibfield  {journal} {\bibinfo  {journal}
  {Phys. Rev.}\ }\textbf {\bibinfo {volume} {D76}},\ \bibinfo {pages} {102004}
  (\bibinfo {year} {2007})}\BibitemShut {NoStop}%
\bibitem [{\citenamefont {Abbott}\ \emph
  {et~al.}(2009{\natexlab{a}})\citenamefont {Abbott} \emph
  {et~al.}}]{Abbott:2009tt}%
  \BibitemOpen
  \bibfield  {author} {\bibinfo {author} {\bibfnamefont {B.~P.}\ \bibnamefont
  {Abbott}} \emph {et~al.} (\bibinfo {collaboration} {LIGO Scientific
  Collaboration}),\ }\href {\doibase 10.1103/PhysRevD.79.122001} {\bibfield
  {journal} {\bibinfo  {journal} {Phys. Rev.}\ }\textbf {\bibinfo {volume}
  {D79}},\ \bibinfo {pages} {122001} (\bibinfo {year} {2009}{\natexlab{a}})},\
  \Eprint {http://arxiv.org/abs/0901.0302} {arXiv:0901.0302 [gr-qc]}
  \BibitemShut {NoStop}%
\bibitem [{\citenamefont {Abbott}\ \emph
  {et~al.}(2009{\natexlab{b}})\citenamefont {Abbott} \emph
  {et~al.}}]{Abbott:2009qj}%
  \BibitemOpen
  \bibfield  {author} {\bibinfo {author} {\bibfnamefont {B.~P.}\ \bibnamefont
  {Abbott}} \emph {et~al.} (\bibinfo {collaboration} {LIGO Scientific
  Collaboration}),\ }\href {\doibase 10.1103/PhysRevD.80.047101} {\bibfield
  {journal} {\bibinfo  {journal} {Phys. Rev.}\ }\textbf {\bibinfo {volume}
  {D80}},\ \bibinfo {pages} {047101} (\bibinfo {year} {2009}{\natexlab{b}})},\
  \Eprint {http://arxiv.org/abs/0905.3710} {arXiv:0905.3710 [gr-qc]}
  \BibitemShut {NoStop}%
\bibitem [{\citenamefont {Abadie}\ \emph
  {et~al.}(2010{\natexlab{b}})\citenamefont {Abadie} \emph
  {et~al.}}]{Abadie:2010yba}%
  \BibitemOpen
  \bibfield  {author} {\bibinfo {author} {\bibfnamefont {J.}~\bibnamefont
  {Abadie}} \emph {et~al.} (\bibinfo {collaboration} {LIGO and Virgo Scientific
  Collaborations}),\ }\href {\doibase 10.1103/PhysRevD.82.102001} {\bibfield
  {journal} {\bibinfo  {journal} {Phys. Rev.}\ }\textbf {\bibinfo {volume}
  {D82}},\ \bibinfo {pages} {102001} (\bibinfo {year} {2010}{\natexlab{b}})},\
  \Eprint {http://arxiv.org/abs/1005.4655} {arXiv:1005.4655 [gr-qc]}
  \BibitemShut {NoStop}%
\bibitem [{\citenamefont {Buonanno}\ \emph {et~al.}(2003)\citenamefont
  {Buonanno}, \citenamefont {Chen},\ and\ \citenamefont {Vallisneri}}]{BCV03b}%
  \BibitemOpen
  \bibfield  {author} {\bibinfo {author} {\bibfnamefont {A.}~\bibnamefont
  {Buonanno}}, \bibinfo {author} {\bibfnamefont {Y.}~\bibnamefont {Chen}}, \
  and\ \bibinfo {author} {\bibfnamefont {M.}~\bibnamefont {Vallisneri}},\
  }\href@noop {} {\bibfield  {journal} {\bibinfo  {journal} {Phys. Rev. D}\
  }\textbf {\bibinfo {volume} {67}},\ \bibinfo {pages} {104025} (\bibinfo
  {year} {2003})},\ \bibinfo {note} {erratum-ibid. {\bf D}~74, 029904(E)
  (2006)},\ \Eprint {http://arxiv.org/abs/gr-qc/0211087} {gr-qc/0211087}
  \BibitemShut {NoStop}%
\bibitem [{lal()}]{lalsuite}%
  \BibitemOpen
  \href@noop {} {\enquote {\bibinfo {title} {L{SC} {A}lgorithm {L}ibrary
  {S}uite},}\ }\bibinfo {howpublished}
  {\url{https://www.lsc-group.phys.uwm.edu/daswg/projects/lalsuite.html}}\BibitemShut
  {NoStop}%
\bibitem [{\citenamefont {Abbott}\ \emph {et~al.}(2010)\citenamefont {Abbott}
  \emph {et~al.}}]{aLIGOSensCurves}%
  \BibitemOpen
  \bibfield  {author} {\bibinfo {author} {\bibfnamefont {B.}~\bibnamefont
  {Abbott}} \emph {et~al.} (\bibinfo {collaboration} {{LIGO Scientific
  Collaboration}}),\ }\href@noop {} {\emph {\bibinfo {title} {Advanced LIGO
  anticipated sensitivity curves}}},\ \bibinfo {type} {Tech. Rep.}\ \bibinfo
  {number} {{LIGO}-T0900288-v3}\ (\bibinfo {year} {2010})\BibitemShut {NoStop}%
\bibitem [{\citenamefont {Poisson}\ and\ \citenamefont {Will}(1995)}]{PW95}%
  \BibitemOpen
  \bibfield  {author} {\bibinfo {author} {\bibfnamefont {E.}~\bibnamefont
  {Poisson}}\ and\ \bibinfo {author} {\bibfnamefont {C.~M.}\ \bibnamefont
  {Will}},\ }\href@noop {} {\bibfield  {journal} {\bibinfo  {journal} {Phys.
  Rev. D}\ }\textbf {\bibinfo {volume} {52}},\ \bibinfo {pages} {848} (\bibinfo
  {year} {1995})}\BibitemShut {NoStop}%
\bibitem [{\citenamefont {Buonanno}\ \emph {et~al.}(2009)\citenamefont
  {Buonanno}, \citenamefont {Iyer}, \citenamefont {Ochsner}, \citenamefont
  {Pan},\ and\ \citenamefont {Sathyaprakash}}]{Buonanno:2009zt}%
  \BibitemOpen
  \bibfield  {author} {\bibinfo {author} {\bibfnamefont {A.}~\bibnamefont
  {Buonanno}}, \bibinfo {author} {\bibfnamefont {B.~R.}\ \bibnamefont {Iyer}},
  \bibinfo {author} {\bibfnamefont {E.}~\bibnamefont {Ochsner}}, \bibinfo
  {author} {\bibfnamefont {Y.}~\bibnamefont {Pan}}, \ and\ \bibinfo {author}
  {\bibfnamefont {B.~S.}\ \bibnamefont {Sathyaprakash}},\ }\href {\doibase
  10.1103/PhysRevD.80.084043} {\bibfield  {journal} {\bibinfo  {journal} {Phys.
  Rev.}\ }\textbf {\bibinfo {volume} {D80}},\ \bibinfo {pages} {084043}
  (\bibinfo {year} {2009})},\ \Eprint {http://arxiv.org/abs/0907.0700}
  {arXiv:0907.0700 [gr-qc]} \BibitemShut {NoStop}%
\bibitem [{\citenamefont {Mikoczi}\ \emph {et~al.}(2005)\citenamefont
  {Mikoczi}, \citenamefont {Vasuth},\ and\ \citenamefont
  {Gergely}}]{Mikoczi:2005dn}%
  \BibitemOpen
  \bibfield  {author} {\bibinfo {author} {\bibfnamefont {B.}~\bibnamefont
  {Mikoczi}}, \bibinfo {author} {\bibfnamefont {M.}~\bibnamefont {Vasuth}}, \
  and\ \bibinfo {author} {\bibfnamefont {L.~A.}\ \bibnamefont {Gergely}},\
  }\href {\doibase 10.1103/PhysRevD.71.124043} {\bibfield  {journal} {\bibinfo
  {journal} {Phys.Rev.}\ }\textbf {\bibinfo {volume} {D71}},\ \bibinfo {pages}
  {124043} (\bibinfo {year} {2005})},\ \Eprint
  {http://arxiv.org/abs/astro-ph/0504538} {arXiv:astro-ph/0504538 [astro-ph]}
  \BibitemShut {NoStop}%
\bibitem [{\citenamefont {Arun}\ \emph {et~al.}(2009)\citenamefont {Arun},
  \citenamefont {Buonanno}, \citenamefont {Faye},\ and\ \citenamefont
  {Ochsner}}]{Arun:2008kb}%
  \BibitemOpen
  \bibfield  {author} {\bibinfo {author} {\bibfnamefont {K.~G.}\ \bibnamefont
  {Arun}}, \bibinfo {author} {\bibfnamefont {A.}~\bibnamefont {Buonanno}},
  \bibinfo {author} {\bibfnamefont {G.}~\bibnamefont {Faye}}, \ and\ \bibinfo
  {author} {\bibfnamefont {E.}~\bibnamefont {Ochsner}},\ }\href {\doibase
  10.1103/PhysRevD.79.104023} {\bibfield  {journal} {\bibinfo  {journal} {Phys.
  Rev.}\ }\textbf {\bibinfo {volume} {D79}},\ \bibinfo {pages} {104023}
  (\bibinfo {year} {2009})},\ \Eprint {http://arxiv.org/abs/0810.5336}
  {arXiv:0810.5336 [gr-qc]} \BibitemShut {NoStop}%
\bibitem [{\citenamefont {Owen}(1996)}]{Owen96}%
  \BibitemOpen
  \bibfield  {author} {\bibinfo {author} {\bibfnamefont {B.~J.}\ \bibnamefont
  {Owen}},\ }\href@noop {} {\bibfield  {journal} {\bibinfo  {journal} {Phys.
  Rev.}\ }\textbf {\bibinfo {volume} {D53}},\ \bibinfo {pages} {6749} (\bibinfo
  {year} {1996})}\BibitemShut {NoStop}%
\bibitem [{\citenamefont {Pai}\ and\ \citenamefont {Arun}(2012)}]{Pai:2012mv}%
  \BibitemOpen
  \bibfield  {author} {\bibinfo {author} {\bibfnamefont {A.}~\bibnamefont
  {Pai}}\ and\ \bibinfo {author} {\bibfnamefont {K.}~\bibnamefont {Arun}},\
  }\href@noop {} {\  (\bibinfo {year} {2012})},\ \Eprint
  {http://arxiv.org/abs/1207.1943} {arXiv:1207.1943 [gr-qc]} \BibitemShut
  {NoStop}%
\bibitem [{\citenamefont {Conway}\ and\ \citenamefont
  {Sloane}(1993)}]{Conway:1993}%
  \BibitemOpen
  \bibfield  {author} {\bibinfo {author} {\bibfnamefont {J.}~\bibnamefont
  {Conway}}\ and\ \bibinfo {author} {\bibfnamefont {N.}~\bibnamefont
  {Sloane}},\ }\href@noop {} {\emph {\bibinfo {title} {Sphere Packings,
  Lattices and Groups}}},\ \bibinfo {edition} {2nd}\ ed.\ (\bibinfo
  {publisher} {Springer-Verlag, New York},\ \bibinfo {year} {1993})\BibitemShut
  {NoStop}%
\bibitem [{\citenamefont {Transtrum}\ \emph {et~al.}(2010)\citenamefont
  {Transtrum}, \citenamefont {Machta},\ and\ \citenamefont
  {Sethna}}]{Transtrum:2010zz}%
  \BibitemOpen
  \bibfield  {author} {\bibinfo {author} {\bibfnamefont {M.~K.}\ \bibnamefont
  {Transtrum}}, \bibinfo {author} {\bibfnamefont {B.~B.}\ \bibnamefont
  {Machta}}, \ and\ \bibinfo {author} {\bibfnamefont {J.~P.}\ \bibnamefont
  {Sethna}},\ }\href {\doibase 10.1103/PhysRevLett.104.060201} {\bibfield
  {journal} {\bibinfo  {journal} {Phys.Rev.Lett.}\ }\textbf {\bibinfo {volume}
  {104}},\ \bibinfo {pages} {060201} (\bibinfo {year} {2010})},\ \Eprint
  {http://arxiv.org/abs/0909.3884} {arXiv:0909.3884 [cond-mat.stat-mech]}
  \BibitemShut {NoStop}%
\bibitem [{\citenamefont {Babak}(2008)}]{Babak:2008rb}%
  \BibitemOpen
  \bibfield  {author} {\bibinfo {author} {\bibfnamefont {S.}~\bibnamefont
  {Babak}},\ }\href {\doibase 10.1088/0264-9381/25/19/195011} {\bibfield
  {journal} {\bibinfo  {journal} {Class. Quant. Grav.}\ }\textbf {\bibinfo
  {volume} {25}},\ \bibinfo {pages} {195011} (\bibinfo {year} {2008})},\
  \Eprint {http://arxiv.org/abs/0801.4070} {arXiv:0801.4070 [gr-qc]}
  \BibitemShut {NoStop}%
\bibitem [{\citenamefont {Manca}\ and\ \citenamefont
  {Vallisneri}(2010)}]{Manca:2009xw}%
  \BibitemOpen
  \bibfield  {author} {\bibinfo {author} {\bibfnamefont {G.~M.}\ \bibnamefont
  {Manca}}\ and\ \bibinfo {author} {\bibfnamefont {M.}~\bibnamefont
  {Vallisneri}},\ }\href {\doibase 10.1103/PhysRevD.81.024004} {\bibfield
  {journal} {\bibinfo  {journal} {Phys.Rev.}\ }\textbf {\bibinfo {volume}
  {D81}},\ \bibinfo {pages} {024004} (\bibinfo {year} {2010})},\ \Eprint
  {http://arxiv.org/abs/0909.0563} {arXiv:0909.0563 [gr-qc]} \BibitemShut
  {NoStop}%
\bibitem [{\citenamefont {Brown}\ \emph {et~al.}(2012)\citenamefont {Brown},
  \citenamefont {Lundgren},\ and\ \citenamefont
  {O'Shaughnessy}}]{Brown:2012gs}%
  \BibitemOpen
  \bibfield  {author} {\bibinfo {author} {\bibfnamefont {D.~A.}\ \bibnamefont
  {Brown}}, \bibinfo {author} {\bibfnamefont {A.}~\bibnamefont {Lundgren}}, \
  and\ \bibinfo {author} {\bibfnamefont {R.}~\bibnamefont {O'Shaughnessy}},\
  }\href@noop {} {\  (\bibinfo {year} {2012})},\ \Eprint
  {http://arxiv.org/abs/arXiv:1203.6060} {arXiv:1203.6060} \BibitemShut
  {NoStop}%
\bibitem [{\citenamefont {Allen}(2005)}]{Allen:2004gu}%
  \BibitemOpen
  \bibfield  {author} {\bibinfo {author} {\bibfnamefont {B.}~\bibnamefont
  {Allen}},\ }\href {\doibase 10.1103/PhysRevD.71.062001} {\bibfield  {journal}
  {\bibinfo  {journal} {Phys. Rev.}\ }\textbf {\bibinfo {volume} {D71}},\
  \bibinfo {pages} {062001} (\bibinfo {year} {2005})},\ \Eprint
  {http://arxiv.org/abs/gr-qc/0405045} {arXiv:gr-qc/0405045} \BibitemShut
  {NoStop}%
\bibitem [{\citenamefont {Robinson}\ \emph {et~al.}(2008)\citenamefont
  {Robinson}, \citenamefont {Sathyaprakash},\ and\ \citenamefont
  {Sengupta}}]{Robinson:2008}%
  \BibitemOpen
  \bibfield  {author} {\bibinfo {author} {\bibfnamefont {C.~A.~K.}\
  \bibnamefont {Robinson}}, \bibinfo {author} {\bibfnamefont {B.~S.}\
  \bibnamefont {Sathyaprakash}}, \ and\ \bibinfo {author} {\bibfnamefont
  {A.~S.}\ \bibnamefont {Sengupta}},\ }\href {\doibase
  10.1103/PhysRevD.78.062002} {\bibfield  {journal} {\bibinfo  {journal} {Phys.
  Rev.}\ }\textbf {\bibinfo {volume} {D78}},\ \bibinfo {pages} {062002}
  (\bibinfo {year} {2008})},\ \Eprint {http://arxiv.org/abs/0804.4816}
  {arXiv:0804.4816 [gr-qc]} \BibitemShut {NoStop}%
\end{thebibliography}%

\end{document}